\documentclass[aps,twocolumn,showpacs,groupedaddress,nofootinbib]{revtex4-1}
\usepackage{graphicx}
\usepackage{amsmath}
\usepackage{amssymb}
\usepackage{color}

\begin{document}
\title{Nonlinear amplification of instabilities with longitudinal expansion}
\author{J\"{u}rgen Berges$^{1,2}$, Kirill Boguslavski$^3$, S\"{o}ren Schlichting$^{1,3}$ \\
$^1$Institut f{\"u}r Theoretische Physik\index{\footnote{}},
Universit{\"a}t Heidelberg\\
Philosophenweg 16, 69120 Heidelberg\\
$^2$ExtreMe Matter Institute EMMI,\\
GSI Helmholtzzentrum\\ 
Planckstr.~1, 64291~Darmstadt\\
$^3$Theoriezentrum, Institut f{\"u}r Kernphysik\\
Technische Universit{\"a}t Darmstadt\\
Schlossgartenstr.~9, 64289 Darmstadt}

\begin{abstract}
We study the dynamics of nonequilibrium instabilities in anisotropically expanding systems. The most prominent example of such a system is the 'Glasma' in the context of relativistic heavy-ion collision experiments, where the expansion is a consequence of approximately boost-invariant initial conditions. Here we consider the problem of parametric resonance in scalar $N$-component quantum field theories with boost-invariant initial conditions, which is similar in spirit. We find that many aspects of the dynamics can be treated analytically by introducing a generalized conformal time. Primary instabilities, which are described by the linearized evolution equations, are seen to lead to a secondary regime of amplifications with strongly enhanced growth rates due to nonlinear corrections. For the secondary instabilities we present a power-counting scheme for weak coupling, and discuss their role for the question of isotropization and the establishment of an equation of state.
\end{abstract}

\maketitle

\section{Introduction}

Plasma instabilities in quantum chromodynamics (QCD) can play an important role for our understanding
of the early time, and maybe even rather late time, evolution of matter created in ultrarelativistic heavy-ion collisions at sufficiently high energies~\cite{Kurkela:2011ub}. Quantitative studies of the evolution at early times are typically performed in either numerical simulations of the classical-statistical field theory \cite{Romatschke:2005pm,Berges:2007re,Kunihiro:2010tg,Fukushima:2011nq} or in simulations of semi-classical transport approaches such as the hard-loop framework~\cite{HL2} and taking into account backreactions on the momentum distribution of the hard particles~\cite{HL3}. 

A crucial open question in this context concerns the role of quantum
fluctuations during the early far-from-equilibrium evolution of the collision. While the origin of an instability can typically be understood from unstable modes of the linearized evolution equations, the inclusion of nonlinear quantum corrections can lead to important phenomena. Most strikingly, the primary growth of fluctuations from unstable modes can lead to a secondary stage of strongly enhanced amplification in a wide momentum range. This can have important consequences for the question of isotropization and the establishment of an equation of state in the context of heavy-ion collisions.

The phenomenon of secondary amplification from nonlinear quantum corrections has been studied in great detail for scalar quantum field theories in the context of early-universe inflaton preheating~\cite{Berges:2002cz,Berges:2004yj}. In that case the dynamics with isotropic expansion can be mapped onto an equivalent problem in Minkowski space-time with the introduction of a conformal time for massless theories. Nonlinear amplifications with strongly enhanced growth rates have also been observed in fixed-box studies of plasma instabilities in $SU(2)$ and $SU(3)$ Yang-Mills theory \cite{Berges:2007re,Fujii:2009kb}, where secondary growth-rates have been pointed out as a possible mechanism to speed-up isotropization. Similar phenomena may also be identified taking into account the longitudinal expansion associated with approximately boost-invariant initial conditions in classical simulations~\cite{Romatschke:2005pm,Fukushima:2011nq}. Here the longitudinal expansion of the plasma leads to a dilution of the system which naturally competes with the instability. Hence it is non-trivial whether primary instabilities can produce high enough occupation numbers for secondary processes to be relevant.

In this work we investigate generic phenomena that occur in anisotropically expanding systems undergoing an instability. We consider scalar $N$-component field theories with parametric resonance initial conditions, for which many aspects can be understood analytically from the underlying quantum field theory by introducing a generalized conformal time. Primary instabilities, which are described by the linearized evolution equations, are described in Sec.~\ref{sec:linearregime}. An important new ingredient going beyond this linear regime is the appearance of the time-dependent spectral function as an additional linearly independent correlation function in the evolution equations. In Sec.~\ref{sec:nonlinear} we present an analytic discussion of the nonlinear amplification of instabilities, where the primary growth of fluctuations is seen to lead to secondary instabilities with strongly enhanced growth rates. We present a power-counting for weak coupling, which takes into account the time dependence of the nonlinear corrections. 

The nonlinear amplification is followed by a nonperturbative regime, where no power-counting for weak coupling can be given. For the scalar theory this regime can be described within a $1/N$ expansion of the two-particle irreducible (2PI) effective action \cite{2PI} to next-to-leading order \cite{JB:2PI1/N}. The dynamics beyond the linear regime can also be described using classical-statistical simulations, which are also available for gauge theories~\cite{Romatschke:2005pm,Berges:2007re,Kunihiro:2010tg,Fukushima:2011nq}.
For comparison with the analytic estimates, we employ a classical-statistical approach for our numerical simulations in Sec.~\ref{sec:classicalstatistical}. We conclude with Sec.~\ref{sec:conclusion}. Further details about the generalized conformal time in anisotropically expanding coordinates as well as the analytic solutions are given in appendices A to D.

\section{Parametric resonance with anisotropic expansion}
\label{sec:linearregime}

\subsection{Scalar field theory and initial conditions}

We consider a real-valued, $N$-component scalar quantum field theory with quartic self-interaction. The classical action is given by
\begin{equation}
S =\! \int\! d^4x \sqrt{-g} \left[\frac{g^{\mu\nu}}{2} \partial_\mu \varphi_a \partial_\nu \varphi_a -\frac{m^2}{2} \varphi_a\varphi_a -\frac{\lambda (\varphi_a\varphi_a)^2}{4!N}\right]  
\label{eq:action}
\end{equation}
with self-coupling $\lambda$ for the fields $\varphi_a(x)$ depending on space-time coordinates $x^\mu = (t,{\bf x})$ and components $a=1,\ldots,N$. Here $g_{\mu\nu}(x)$ is the metric tensor and its determinant is denoted by $g(x)=\det g_{\mu\nu}(x)$. Since boost-invariant initial conditions along the longitudinal direction $z \equiv x^3$ will be of interest, it is convenient to introduce the comoving coordinates
\begin{eqnarray}
\tau=\sqrt{t^2-z^2} \;,\qquad \eta=\text{artanh}\left(\frac{z}{t}\right)  \;.
\label{eq:coord}
\end{eqnarray}
The fields becomes then a function of 'proper time' $\tau$, rapidity $\eta$ and the remaining transverse coordinates $x_T = (x^1, x^2)$. The metric tensor in comoving coordinates follows from the transformation from Minkowski space as $g_{\mu\nu}=\mathrm{diag}(1,-1,-1,-\tau^2)$ and $\sqrt{-g}=\tau$. 

We are interested in the real-time evolution of the corresponding quantum field theory for scalar $N$-component Heisenberg field operators $\hat{\phi}_a(\tau,x_T,\eta)$.
For a given density matrix $\rho_D(\tau_0)$ at initial proper time $\tau_0$ the field expectation value is $\langle \hat{\phi}(\tau,x_T,\eta) \rangle \equiv \text{Tr}[\rho_D(\tau_0) \hat{\phi}(\tau,x_T,\eta)]$ and equivalently for products of field operators, which determine the correlation functions. We will consider $\tau_0>0$, i.e. the time-evolution in the forward light-cone, and spatially homogeneous expectation values. By spatially homogeneous we mean homogeneous in the transverse plane and homogeneous in longitudinal rapidity at given proper time.
We employ Gaussian initial conditions, which can be conveniently formulated in terms of the macroscopic field 
\begin{equation}
\phi_a(\tau) \, = \, \left\langle \hat{\phi}_a(\tau,x_T,\eta)\right\rangle \, 
\label{eq:deffield}
\end{equation}
as well as the statistical two-point correlation function
\begin{eqnarray}
&& F_{ab}(\tau,\tau',x_T- x_T',\eta-\eta') \nonumber\\ 
&& = \frac{1}{2}\left\langle\left\{\hat{\phi}_a(\tau,x_T,\eta),\hat{\phi}_b(\tau',x_T',\eta')\right\}\right\rangle
-\phi_a(\tau) \phi_b(\tau') 
\label{eq:defF}
\end{eqnarray}
and derivatives at initial time. Here $\{.,.\}$ denotes the anti-commutator. Using $O(N)$ symmetry in field-index space, we take the macroscopic field to point in the $a=1$ direction, i.e.\ 
\begin{equation}
\phi_a(\tau) = \phi(\tau) \delta_{a1} \, .
\label{eq:fielddirection}
\end{equation} 
The fluctuations can be taken to be diagonal, i.e.\ $F_{ab}=\text{diag}(F_{\|},F_{\bot},\ldots,F_{\bot})$ with the subscripts, longitudinal and transverse, indicating the orientation with respect to the macroscopic field.

Relevant for parametric resonance are weak couplings ($\lambda \ll 1$) with a parametrically large macroscopic field ($\phi^2(\tau_0) \sim {\cal{O}}(1/\lambda)$) and small fluctuations ($F(\tau_0) \sim {\cal{O}} (1)$) at initial time. For the field initial conditions we write
\begin{eqnarray}
\phi(\tau_0)&=&\sqrt{\frac{6N}{\lambda}}\, \sigma_0\,, \nonumber \\
\left.\partial_{\tau}\phi(\tau)\right|_{\tau=\tau_0}&=&-\frac{1}{3\tau_0}\, \phi(\tau_0) \, ,
\label{eq:phiIC}
\end{eqnarray} 
where the derivative at $\tau=\tau_0$ mimics an initially free-streaming behavior and $\sigma_0$ denotes the rescaled initial field amplitude. If not stated otherwise we will employ the typical set of parameters $m^2=0$, $\sigma_0\tau_0=5$, $N=4$ and $\lambda=10^{-4}$ for our estimates and numerical simulations below. The spectral shape of the initial correlation functions will be of little relevance as long as the magnitude is sufficiently small compared to the initial macroscopic field squared. For completeness the spectral distribution of the considered fluctuations is given in appendix D.

\subsection{Linear regime and generalized conformal time}

Since the initial macroscopic field $\phi$ is large and fluctuations are small, at sufficiently early times the time evolution is accurately described by the classical evolution of the field and linearized fluctuations around it. The subsequent onset of non-linearities and impact of quantum fluctuations will be discussed further below. In the linear regime the evolution of the macroscopic field follows directly from the stationarity of the classical action (\ref{eq:action}) and is given in comoving coordinates by the equation of a damped anharmonic oscillator
\begin{equation}
\left[\partial_{\tau}^2+\frac{1}{\tau}\partial_{\tau}+m^2+\frac{\lambda}{6N}\phi^2(\tau)\right]\phi(\tau) \, = \, 0 \;.
\label{eq:phiEvoEq} 
\end{equation}
The damping term characterized by a first-order $\tau$-derivative arises due to the longitudinal expansion and represents dilution of the system. In order to solve (\ref{eq:phiEvoEq}) analytically, we perform a change in the time variable as well as a rescaling of the fields by introducing
\begin{equation}
\sigma_0 d\tau \, = \, \left(\frac{a(\tau)}{a(\tau_0)}\right)^{1/3}d\theta\;, \qquad \tilde{\phi}=\left(\frac{a(\tau)}{a(\tau_0)}\right)^{1/3} \frac{\phi}{\sigma_0} 
\end{equation}
with the generalized scale factor
\begin{equation}
 a(\tau) \, = \, \tau 
\end{equation}
in case of a one-dimensional Bjorken expansion. This can be seen as a generalization of the concept of conformal time to anisotropically expanding systems. For the one-dimensional expansion the scale factor $a(\tau)$ enters the above definition of conformal time with exponent $1/3$, whereas in the isotropically expanding case one would use $d/3$ for $d$-dimensional space instead. The new dimensionless time-variable $\theta$ is explicitly given by
\begin{equation}
\theta-\theta_0 \, = \, \frac{3}{2}\, \sigma_0\tau_0 \left[\left(\frac{\tau}{\tau_0}\right)^{2/3}-1\right] 
\label{eq:ctExpression}
\end{equation}
and we will refer to it as conformal time in analogy to the isotropic case. The evolution equation of the macroscopic field in terms of the new variables reads
\begin{equation}
\left[\partial_\theta^2+\tilde{m}^2(\theta)+\frac{\lambda}{6N}\,\tilde{\phi}^2(\theta)\right]\tilde{\phi}(\theta)=0 \;,
\label{eq:macPhiCTeom}
\end{equation}
where we introduced the time-dependent effective 'mass term' 
\begin{equation}
\tilde{m}^2(\theta)=\frac{m^2}{\sigma_0^2} \left(\frac{a(\theta)}{a(\theta_0)}\right)^{2/3}+\frac{2}{9} \left(\frac{a'(\theta)}{a(\theta)}\right)^2-\frac{1}{3}\frac{a''(\theta)}{a(\theta)}\,.
\label{eq:massTildeDef}
\end{equation}
Here primes denote derivatives with respect to $\theta$, which are explicitly given by
\begin{eqnarray}
\frac{a(\theta)}{a(\theta_0)} &=& \left(\frac{2\theta}{3\sigma_0\tau_0}\right)^{3/2} \;,\quad
\frac{a'(\theta)}{a(\theta_0)} \, = \, \frac{1}{\sigma_0\tau_0}\left(\frac{2\theta}{3\sigma_0\tau_0}\right)^{1/2} \;,
\nonumber\\
\frac{a''(\theta)}{a(\theta_0)} &=& \frac{1}{3(\sigma_0\tau_0)^2}\left(\frac{2\theta}{3\sigma_0\tau_0}\right)^{-1/2}\;,
\end{eqnarray}
where we chose $\theta_0=3\sigma_0\tau_0/2$ in accordance with (\ref{eq:ctExpression}). This change of variables is discussed in more detail in appendix A.

\begin{figure}[tp]
\centering
\includegraphics[scale=0.68]{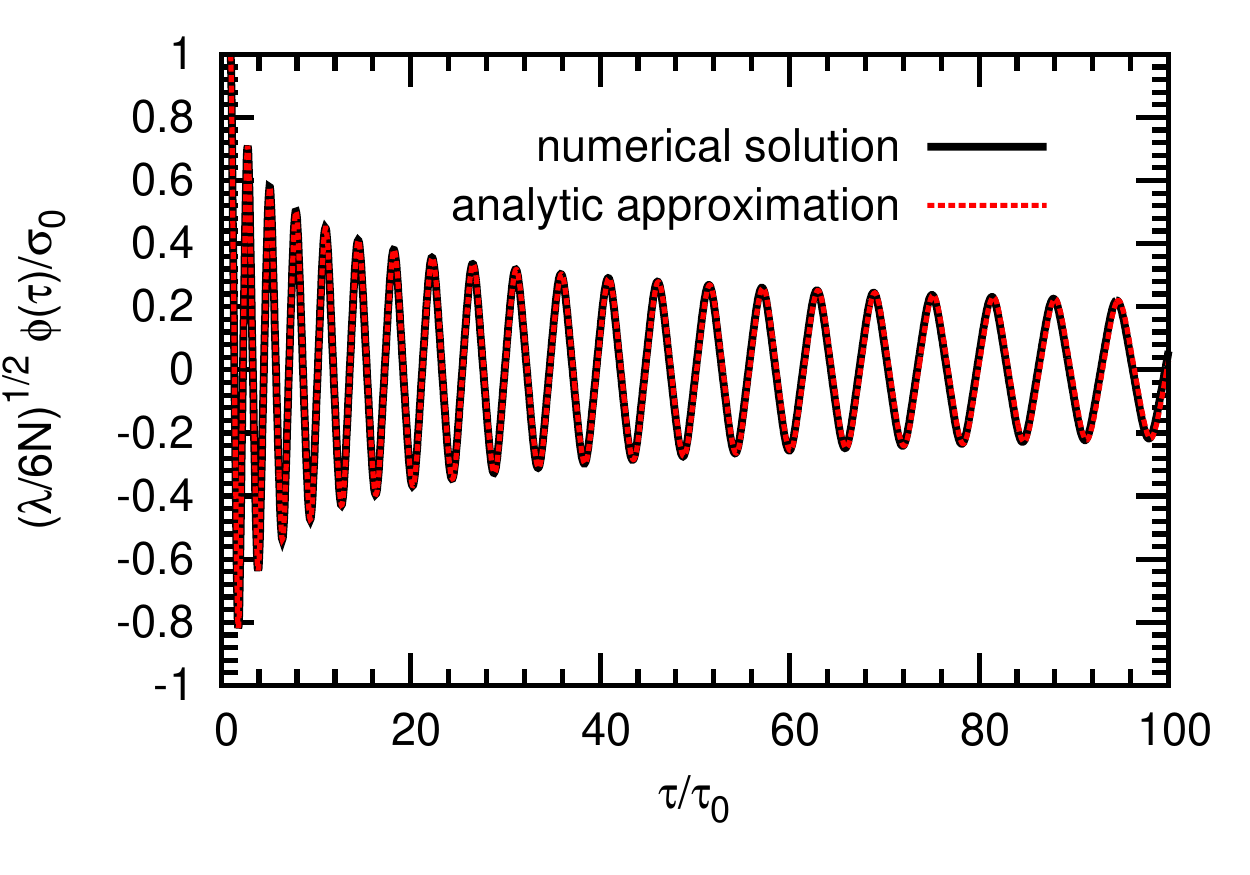}
\caption{\label{fig:linPhi} (color online) Solution to the linearized field equation (\ref{eq:phiEvoEq}) for $\sigma_0\tau_0=5$. The approximate analytical solution (\ref{eq:phiAnaSol}) given by the dashed curve is practically on top of the full numerical one (solid curve). As a consequence of the longitudinal expansion, the field shows oscillations with constant period in $\tau^{2/3}$ while it decreases as $\tau^{-1/3}$.}
\end{figure}

For large initial field amplitude, $\sigma_0\tau_0 \gg 1$, the above evolution equation can be solved approximately for the massless case ($m^2=0$). According to (\ref{eq:massTildeDef}) one then has $\tilde{m}^2(\theta)=1/(4\theta^2)$, which is small compared to the macroscopic field. Thus $\tilde{m}^2(\theta)\simeq 0$ gives a rather accurate description of the dynamics in this case. The evolution equation (\ref{eq:macPhiCTeom}) then becomes that of an anharmonic oscillator
\begin{equation}
\left[\partial_\theta^2+\frac{\lambda}{6N}\tilde{\phi}^2(\theta)\right]\tilde{\phi}(\theta) \, = \, 0 \;.
\label{eq:phiAnaEq}
\end{equation}
Its solution is given in terms of Jacobi elliptic functions, i.e.
\begin{eqnarray}
\tilde{\phi}(\theta)=\sqrt{\frac{6N}{\lambda}}~\text{cn}\left(\theta-\theta_0\, ;\frac{1}{2}\right) 
\label{eq:phiSolTheta}
\end{eqnarray}
for the considered initial conditions specified in (\ref{eq:phiIC}) and appendix D. 
The Jacobi cosine, $\text{cn}(\theta;\alpha)$, is a doubly periodic function in $\theta$ with periods $4K(\alpha)$ and $4iK(1-\alpha)$, where $K(\alpha)$ is the complete elliptic integral of the first kind.\footnote{In terms of a series expansion,  $K(\alpha)=\frac{\pi}{2}+\frac{\pi}{2}\alpha+\frac{9\pi}{128}\alpha^2+\frac{25\pi}{512}\alpha^3+\mathcal{O}(\alpha^4)$.} Accordingly the macroscopic field $\tilde{\phi}$ displays oscillations with constant period in conformal time. For the physical interpretation of this result it is insightful to express the approximate solution (\ref{eq:phiSolTheta}) in terms of the original variables, where it reads
\begin{equation}
\phi(\tau) =  \sigma_0 \sqrt{\frac{6N}{\lambda}}\left(\frac{\tau_0}{\tau}\right)^{1/3} \text{cn}\left(\frac{3\sigma_0 \tau_0}{2} \left[ \left(\frac{\tau}{\tau_0}\right)^{2/3}-1\right];\frac{1}{2}\right).
\label{eq:phiAnaSol}
\end{equation}
In Fig.~\ref{fig:linPhi} this is compared to the numerical solution of (\ref{eq:phiEvoEq}) without further approximations. The very good agreement verifies that the terms neglected in (\ref{eq:phiAnaEq}) are irrelevant for this choice of parameters. 

From Fig.~\ref{fig:linPhi} one observes that the macroscopic field $\phi(\tau)$ decays as $\tau^{-1/3}$, while it displays oscillatory behavior with a constant period in conformal time $\theta\propto\tau^{2/3}$. The prior follows, of course, also from simple considerations of energy-momentum conservation. The energy density $\epsilon$ satisfies the equation of Bjorken hydrodynamics
\begin{eqnarray}
\frac{d \epsilon}{d\tau}=-\frac{\epsilon+P_L}{\tau} \;,
\end{eqnarray}
where $P_L$ denotes the longitudinal pressure in the local rest-frame.\footnote{See Sec.\ 4 for a discussion of the stress-energy tensor.} Initially the energy density is dominated by the macroscopic field and proportional to $\phi^4$. As the macroscopic field is homogeneous and the system is conformally invariant, the longitudinal pressure averages to $1/3$ of the energy density. Hence $\epsilon \sim \tau^{-4/3}$ and, therefore, $\phi \sim \tau^{-1/3}$. The fact that the oscillations have a constant period in conformal time is a consequence of the dilution of the system. 

\subsection{Nonequilibrium instability with expansion}
\label{sec:instexp}

To describe the linearized evolution equations for the fluctuations (\ref{eq:defF}), we will work in Fourier space with respect to spatial coordinates, i.e.\ we consider the dimensionless quantity
\begin{equation}
F_{\bot}(\tau,\tau',\vec{p}_T,\nu)=\int d^2 x_T d\eta\, F_{\bot}(\tau,\tau',\vec{x}_T,\eta)\, e^{-i(\vec{p}_T\vec{x}_T+\nu\eta)}\; 
\end{equation}
and equivalently for the longitudinal component $F_{\|}(\tau,\tau',p_T,\nu)$. The linearized evolution equations for the fluctuations can be obtained by expanding the classical field equation of motion, given by the stationarity of the action (\ref{eq:action}), to first order in deviations from the homogeneous background field $\phi(\tau)$. Later in Sec.~\ref{sec:nonlinear} on quantum corrections, these equations will also be seen to correspond to the evolution equations of the quantum field theory in the limit where all nonlinear or loop corrections are neglected. The linearized equations for the fluctuations read
\begin{eqnarray}
\left[\partial_{\tau}^2+\frac{1}{\tau}\partial_{\tau}+p_T^2+\frac{\nu^2}{\tau^2} + m^2 + \frac{\lambda}{6N}\phi^2(\tau)\right]F_{\bot}(\tau,\tau',p_T,\nu) &&
\nonumber \\
= \, 0 , &&
\nonumber \\
\left[\partial_{\tau}^2+\frac{1}{\tau}\partial_{\tau}+p_T^2+\frac{\nu^2}{\tau^2} + m^2 + \frac{\lambda}{2N}\phi^2(\tau)\right]F_{\|}(\tau,\tau',p_T,\nu) &&
\nonumber \\
= \, 0 . &&
\nonumber\\ 
&&
\label{eq:LinFT} 
\end{eqnarray}
The damping term characterized by a first-order $\tau$-derivative again appears due to the expansion. Other competing scales that depend differently on proper-time $\tau$ are related to
\begin{itemize}
\item the macroscopic field squared $\phi^2(\tau)\sim \tau^{-2/3}$,
\item the longitudinal momentum squared $\nu^2/\tau^2$ and
\item the transverse 'mass' squared $m^2+p_T^2$ which is independent of proper-time.
\end{itemize}
This interplay of dilution and red-shift effects induces a variety of new phenomena, which are not present for non-expanding systems. In particular, the well-known phenomenon of parametric resonance in Minkowski space-time will receive significant changes in the presence of longitudinal expansion. 

For an analytic description of the time evolution of fluctuations,
we introduce conformal time variables similar to what has been done for the macroscopic field above:
\begin{equation}
\tilde{F}(\theta,\theta',p_T,\nu) = \left(\frac{\tau}{\tau_0}\right)^{1/3}\left(\frac{\tau'}{\tau_0}\right)^{1/3}F(\tau,\tau',p_T,\nu),
\end{equation}
where $\tau=\tau(\theta)$ and accordingly $\tau'=\tau(\theta')$. Inserting the solution for the background field (\ref{eq:phiSolTheta}), the linearized evolution equations using conformal time variables read
\begin{eqnarray}
\left[\partial_{\theta}^2 + \tilde{p}^2(\theta) +\tilde{m}^2(\theta)+\text{cn}^2 \left(\theta-\theta_0\, ;\frac{1}{2}\right)\right]\tilde{F}_{\bot}(\theta,\theta',p_T,\nu)  &&
\nonumber \\
= \, 0 , &&
\nonumber \\
\left[\partial_{\theta}^2 + \tilde{p}^2(\theta) +\tilde{m}^2(\theta)+3\, \text{cn}^2 \left( \theta-\theta_0 \, ;\frac{1}{2}\right)\right]\tilde{F}_{\|}(\theta,\theta',p_T,\nu) && 
\nonumber \\
= \, 0 .
\nonumber \\
\label{eq:EOMFtrans}
\end{eqnarray}
We restrict ourselves again to the massless case, $m^2 = 0$, and approximate $\tilde{m}^2(\theta)\simeq 0$ as in (\ref{eq:phiAnaEq}). With this approximation the equation of motion resembles the Jacobian form of the Lam\'{e} equation.  A crucial difference is the explicit $\theta$-time dependence of the (dimensionless) momentum term 
\begin{equation}
\tilde{p}^2(\theta) \, \equiv \, \frac{2\theta}{3\sigma_0\tau_0}\, \frac{p_T^2}{\sigma_0^2} + \frac{9}{4}\frac{\nu^2}{\theta^2} \, .
\label{eq:mt}
\end{equation} 

It is instructive to consider for a moment what happens if $\tilde{p}^2$ was not depending on $\theta$-time. In this case (\ref{eq:EOMFtrans}) corresponds for vanishing $\tilde{m}^2(\theta)$ to Lam\'{e} equations. The latter exhibit the well-known phenomenon of parametric resonance, which is reviewed in appendix B. The dynamics is then dominated by the transverse modes ($\tilde{F}_{\bot}$) showing exponential growth as a function of time in a resonance band of momenta with   
\begin{eqnarray}
0 \, \leq \, \tilde{p}^2 \, \leq \, \frac{1}{2} \, . 
\label{eq:GrowthCond}
\end{eqnarray}
For our purposes the (dimensionless) momentum-dependent rate for exponential growth is to good approximation given by 
\begin{eqnarray}
 \gamma(\tilde{p})\, \simeq \, \frac{4\pi e^{-\pi}}{K(1/2)} \sqrt{2\tilde{p}^{2}\left(1-2\tilde{p}^{2}\right)} \,,
\label{eq:growthrate}
\end{eqnarray}
where the maximum growth rate in this approximation is
\begin{eqnarray}
\gamma_{0}\, \simeq \, \frac{2\pi e^{-\pi}}{K(1/2)}\;,\qquad \text{for} \qquad \tilde{p}_{0}^2 \, \simeq \, \frac{1}{4}\;.
\label{eq:maxgrowth}
\end{eqnarray} 
We give more accurate analytic expressions involving inverse Jacobi elliptic and Jacobi zeta functions in appendix B, where we also compare the approximate expression (\ref{eq:growthrate}) to the full growth rate in Fig.~\ref{fig:grExactVsApprox}.

\begin{figure*}[t]
\includegraphics[scale=0.7]{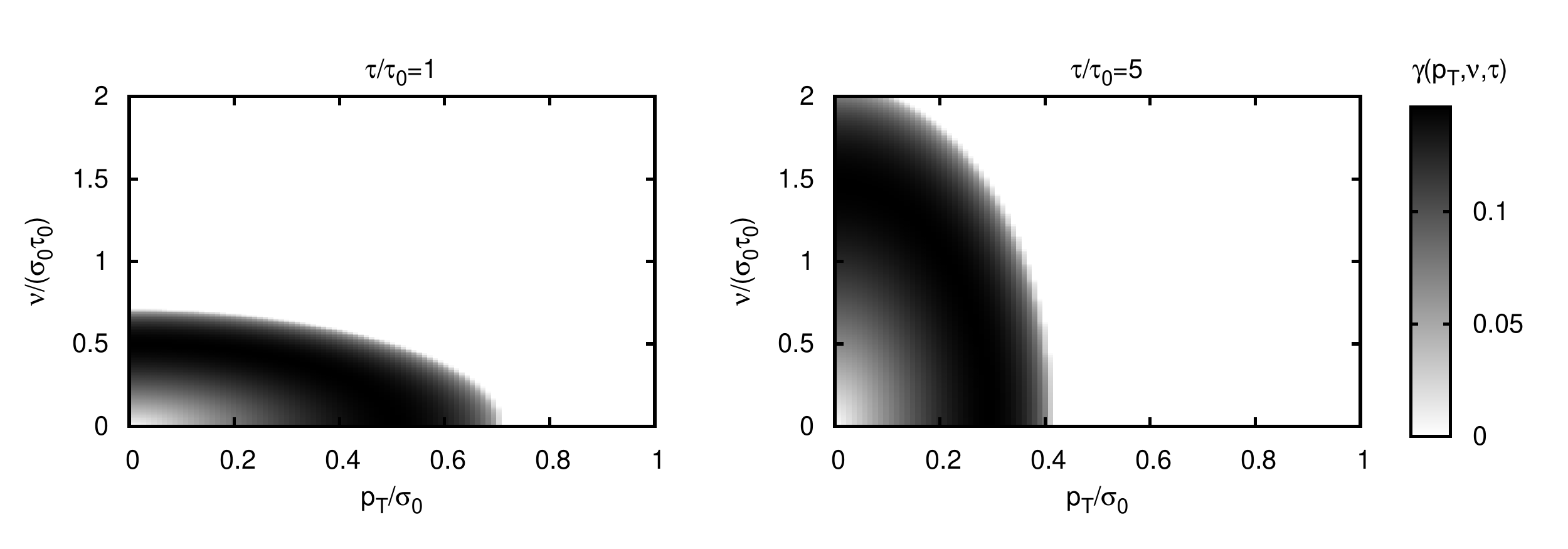}
\caption{\label{fig:tdGR} Time-dependent (dimensionless) growth rate $\gamma(p_T,\nu,\theta)$ in the transverse momentum $p_T$ and rapidity wave number $\nu$ plane for different times. The (\textbf{left}) panel corresponds to $\tau/\tau_0=1$, the (\textbf{right}) panel shows $\tau/\tau_0=5$. The instability develops from the high $p_T$ small $\nu$ region to the high $\nu$ small $p_T$ region.}
\end{figure*}

For the anisotropically expanding system the situation is somewhat more involved due to the effects of dilution and the red-shift of longitudinal momentum modes. In order to obtain an analytic understanding for the expanding system described by (\ref{eq:EOMFtrans}), it is very instructive to assume that the momentum term (\ref{eq:mt}) varies only slowly on the characteristic time scale of one oscillation of the macroscopic field. By expanding the time dependent momentum term as $\tilde{p}^2(\theta+T_{\theta})\simeq\tilde{p}^2(\theta)+\partial_\theta\tilde{p}^2(\theta)T_{\theta}$, where $T_{\theta}=4K(1/2)$ corresponds to one period of oscillation of the macroscopic field, we find that formally this quasi-static approximation corresponds to the limit $4K(1/2)|\partial_\theta \tilde{p}^2(\theta)|\ll \tilde{p}^2(\theta)$. For resonant modes, where the characteristic momenta are parametrically given by $p_T\sim\sigma_0$ and $\nu\sim\sigma_0\tau_0$, it is straightforward to verify that the above condition is approximately fulfilled whenever $\sigma_0\tau_0\gtrsim 4~K(1/2)$, which can easily be achieved by choice of parameters. Within this approximation the resonance band and growth rates for the expanding system are obtained from (\ref{eq:GrowthCond}) and (\ref{eq:growthrate}) by replacing $\tilde{p}^2 \rightarrow \tilde{p}^2(\theta)$ as given in (\ref{eq:mt}). Therefore, in this approximation the time-dependence of the growth rate enters only through the explicit time-dependence of $\tilde{p}^2(\theta)$ and is a consequence of the red-shift and the dilution. 

The growth rate (\ref{eq:growthrate}) with time-dependent momentum terms (\ref{eq:mt}) is displayed graphically in Fig.~\ref{fig:tdGR} as a function of transverse momentum $p_T$ and longitudinal wave number $\nu$ for different times. One observes that the instability develops from the high $p_T$ and small $\nu$ region at early times to the high $\nu$ and small $p_T$ region at later times. In the remainder of this section we will discuss different phenomena that appear within the quasi-static approach and show that these analytic results compare well to full numerical solutions of (\ref{eq:EOMFtrans}). 

The explicit $\theta$-time dependence of the approximate resonance band criterion leads to new phenomena that are not present in the non-expanding case. A given mode with fixed transverse momentum $p_T$ and fixed $\nu$ may satisfy the condition (\ref{eq:GrowthCond}) for a certain time window $\tau_{\text{Start}}<\tau<\tau_{\text{End}}$, while the condition is not met outside this window. Here $\tau_{\text{Start}}$ and $\tau_{\text{End}}$ are related to the corresponding conformal times by (\ref{eq:ctExpression}). This means that modes can shift inside the resonance band, exhibit exponential growth for a certain time and then shift back out of the resonance band again so that the exponential growth stops. In particular, if $\tau_{\text{Start}}>\tau_0$ for a certain mode, this mode will exhibit a delay in the onset-time of growth. We note that this qualitative behavior is characteristic for nonequilibrium instabilities also in other systems. For instance, a similar behavior has been observed in numerical studies of gauge field evolution for plasma instabilities in the context of relativistic heavy-ion collisions \cite{Romatschke:2005pm}, where only the $\nu$ dependence was studied. If we also restrict to the case of vanishing transverse momentum, the condition (\ref{eq:GrowthCond}) yields the onset of exponential growth at times
\begin{eqnarray}
\left(\frac{\tau_{\text{Start}}}{\tau_0}\right)^{2/3} \, = \, \frac{\sqrt{2\nu^2}}{\sigma_0\tau_0} \qquad\text{for} \qquad p_T =0 \, .
\label{eq:tLinSetIn}  
\end{eqnarray}
For these modes the subsequent growth continues as long as the linearized description remains valid. However, by investigating the entire momentum space one finds that there are modes which behave quite differently. Modes with vanishing longitudinal momentum exhibit a resonant amplification only until the time
\begin{eqnarray}
\left(\frac{\tau_{\text{End}}}{\tau_0}\right)^{2/3} \, = \, \frac{\sigma_0^2}{2 p_T^2} \qquad\text{for} \qquad \nu=0.
\label{eq:tLinFO} 
\end{eqnarray}
After this time these modes freeze out and exponential growth stops, while they start showing stable oscillatory behavior. 

The generic situation, where neither longitudinal nor transverse momentum vanish, is discussed in detail in appendix C and we refer here only to some characteristic results.  By searching for real positive solutions for $\theta$ of the resonance criterion (\ref{eq:GrowthCond}) as relevant for the initial value problem, we find that these solutions exist for all modes satisfying the time-independent condition
\begin{eqnarray}
\frac{p_T^4\nu^2}{\sigma_0^6\tau_0^2} \, \leq \, \frac{1}{54} \;.
\label{eq:solGrowthCond}
\end{eqnarray}
Accordingly, all transverse modes ($F_{\bot}$) satisfying the condition (\ref{eq:solGrowthCond}) experience exponential amplification for a certain period of time. The set-in and freeze-out times for these modes correspond to times where the relation (\ref{eq:GrowthCond}) is taken as an equality. Calling $\xi \equiv (54~p_T^4\nu^2)/(\sigma_0^6 \tau_0^2)$ these times are given by 
\begin{eqnarray}
\left(\frac{\tau_{\text{Start}}}{\tau_0}\right)^{2/3}\! & = & \frac{\sigma_0^2}{6 p_T^2} \left[1+2\sin\left(\frac{2}{3}\, \text{arctan}\sqrt{\frac{\xi}{1-\xi}}-\frac{\pi}{6}\right)\right]\! , 
\nonumber\\
\left(\frac{\tau_{\text{End}}}{\tau_0}\right)^{2/3}\! & = & \frac{\sigma_0^2}{6 p_T^2} \left[1+2\cos\left(\frac{2}{3}\, \text{arctan}\sqrt{\frac{\xi}{1-\xi}}\right)\right] \!,
\end{eqnarray}
where $\xi \leq 1$ and, of course, $\tau_{\text{Start}} \leq \tau_{\text{End}}$ for unstable modes according to (\ref{eq:solGrowthCond}). 

\begin{figure}[t]
\includegraphics[scale=0.7]{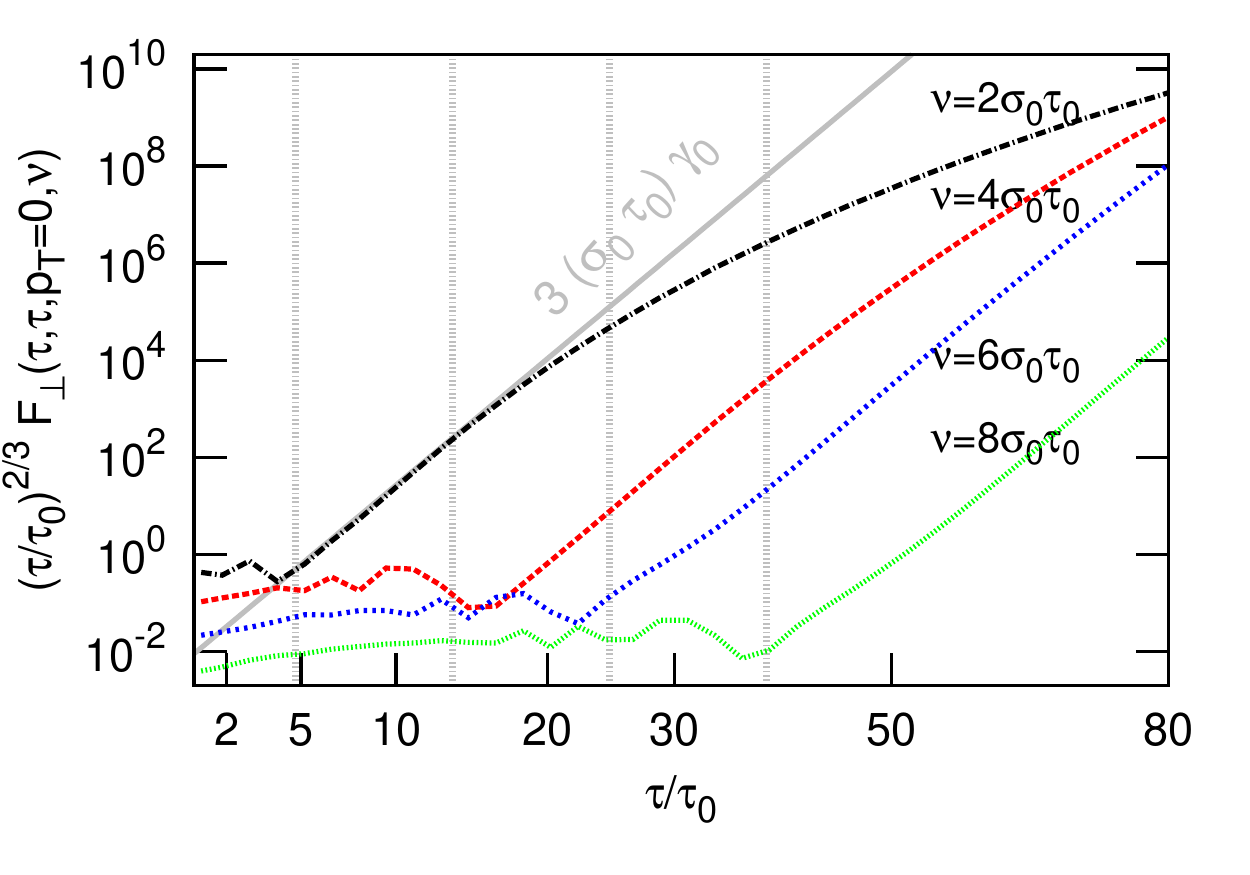}
\caption{ \label{fig:linNUModes} (color online) Time evolution of the equal-time transverse fluctuations $F_{\bot}(\tau,\tau,p_T,\nu)$ from the linearized evolution equations for $p_T=0$ and different rapidity wave numbers $\nu$. The modes are averaged over one period of oscillation of the macroscopic field. The time axis is scaled as $(\tau/\tau_0)^{2/3}$. The vertical dashed grey lines represent the set-in time of the instability according to (\ref{eq:tLinSetIn}). The full grey line corresponds to the maximum growth rate $3 \sigma_0\tau_0 \gamma_0$.}
\end{figure}

\begin{figure}[t]
\centering
\includegraphics[scale=0.7]{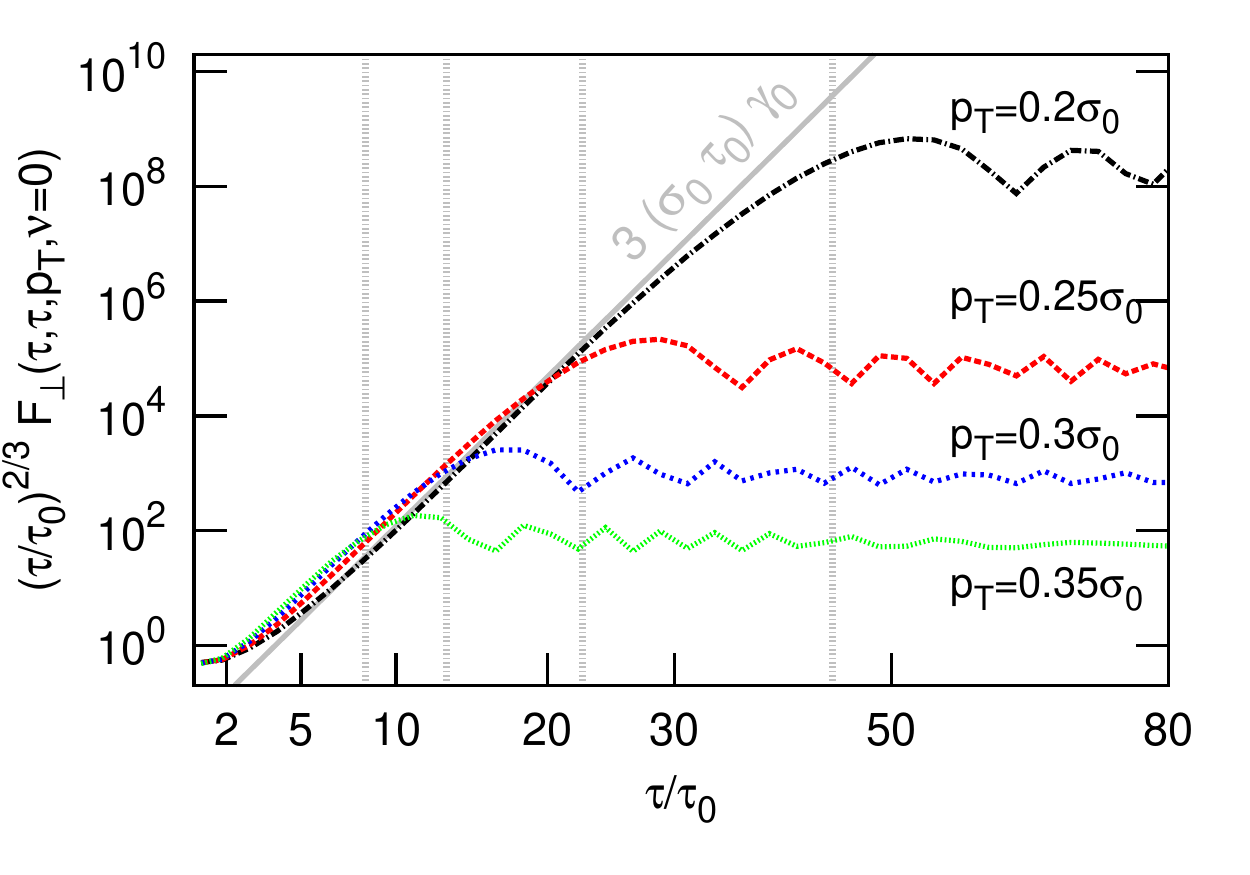}
\caption{\label{fig:linPTModes} (color online) Time evolution of the transverse fluctuations $F_{\bot}(\tau,\tau,p_T,\nu)$ from the linearized evolution equations for $\nu=0$ and different transverse momenta $p_T$. The modes are averaged over one period of oscillation of the macroscopic field. The time axis is scaled as $(\tau/\tau_0)^{2/3}$.  The vertical grey dashed lines correspond to the estimates for the freeze-out times of the instability according to (\ref{eq:tLinFO}). The full grey line corresponds to the maximum growth rate $3 \sigma_0 \tau_0 \gamma_0$.}
\end{figure}

The phenomena of delayed set-in and freeze-out can be observed in our numerical studies of the linearized evolution equations. In Figs.~\ref{fig:linNUModes} and \ref{fig:linPTModes} we present the numerical solution of the linearized evolution equations for transverse modes $F_{\bot}(\tau,\tau,p_T,\nu)$ with different transverse and longitudinal momenta. From Fig.~\ref{fig:linPTModes} one observes that modes which are dominated by their transverse momentum exhibit an amplification at early times. This amplification stops when they shift out of the resonance band. We find that this behavior is indeed very well described by our analytic estimate (\ref{eq:tLinFO}). In contrast, modes which are dominated by their rapidity wave number $\nu$, as shown in Fig.~\ref{fig:linNUModes}, exhibit an amplification at later times, when they have shifted inside the resonance band. Here the set-in time of the instability is well described by (\ref{eq:tLinSetIn}). In both cases we find that the growth is exponential in conformal time $\theta\propto \tau^{2/3}$ but with a time-dependent growth-rate $\gamma(\tilde{p}(\theta))\leq\gamma_0$. 

For instance, if we consider modes with small transverse momentum, such as $p_T=0.2 \sigma_0$ and $\nu=0$ as depicted in Fig.~\ref{fig:linPTModes}, one observes that the growth rate is small initially, then closely approaches the maximum value of $3 \sigma_0 \tau_0 \gamma_0$ and subsequently decreases until freeze-out occurs. The maximum growth rate\footnote{The maximum growth rate for equal-time correlation functions $\tilde{F}(\theta,\theta,p_T,\nu)$ is given by $2 \gamma_0$. In terms of the proper-time variable $\tau$ this translates into $3 \sigma_0 \tau_0 \gamma_0$.}  $3 \sigma_0 \tau_0 \gamma_0$ is realized only when $\tilde{p}^2(\theta) = \tilde{p}^2_0$, where $\tilde{p}_0$ is the maximally amplified momentum given in (\ref{eq:maxgrowth}). Hence if $\tilde{p}^2(\theta_0) < \tilde{p}^2_0$ initially, the modes experience a smaller growth rate at early times. For the considered mode $\tilde{p}^2(\theta)\sim \theta \, p_T^2$ and, therefore, it increases with (conformal) time. Hence $\gamma(\tilde{p}(\theta))$ increases until it reaches the maximum rate when $\tilde{p}(\theta)=\tilde{p}_0$ and subsequently decreases until the mode shifts out of the resonance band and the freeze-out occurs. This can be thought of as traversing the growth-rate $\gamma(\tilde{p})$ -- as shown graphically in Fig.~\ref{fig:grExactVsApprox} in appendix B -- from smaller to higher momenta (left to right in Fig.~\ref{fig:grExactVsApprox}) in time. In contrast, modes with $p_T=0$ and non-vanishing $\nu$, as shown in Fig.~\ref{fig:linNUModes}, are characterized by a $\tilde{p}(\theta)$ which decreases with time. Therefore, one finds that the largest growth rate is realized shortly after set-in of the instability and decreases towards later times. This corresponds to traversing the growth-rate $\gamma(\tilde{p})$ in the opposite direction, i.e.\ from higher to smaller momenta (right to left in Fig.~\ref{fig:grExactVsApprox}) in time.

\section{Nonlinear amplification of instabilities}
\label{sec:nonlinear}

\subsection{Nonlinear corrections and power counting}

How to take into account nonlinear corrections in scalar quantum field theories is well known and can be based efficiently on the two-particle irreducible (2PI) effective action \cite{2PI}, which circumvents problems of secular time evolutions encountered in non-resummed (1PI) approximation schemes \cite{Berges:2004yj}. For the $N$-component scalar quantum field theory a non-perturbative description can be based on the $1/N$-expansion of the 2PI effective action to next-to-leading order (NLO), which is explained in Refs.~\cite{JB:2PI1/N}. The evolution equations have been presented for general metric in Refs.~\cite{Tranberg,Stefan}. Gauge theories with expansion have also been considered in Ref.~\cite{Hatta:2PICGC} based on a classification of the 2PI effective action in the number of loops.

Here we discuss those aspects of the time evolution, which are relevant for an understanding of the nonlinear amplification of instabilities. Nonlinear corrections will lead to strongly enhanced 'secondary' growth rates, which are multiples of the initial 'primary' growth rates observed from the linear regime in Sec.~\ref{sec:linearregime}. Remarkably, this turns out to be very similar to the non-expanding case, which has been discussed extensively in the past for scalars \cite{Berges:2002cz,Berges:2004yj} as well as pure gauge theories \cite{Berges:2007re}.  

The linear regime discussed above is described in terms of the macroscopic or background field $\phi$, defined in (\ref{eq:deffield}), and the fluctuation or statistical two-point function  $F$ as given by (\ref{eq:defF}). Together, the corresponding evolution equations (\ref{eq:phiEvoEq}) and (\ref{eq:LinFT}) form a closed set of equations for the linear regime. An important new ingredient going beyond the linear regime will be the appearance of the time-dependent spectral function as an additional linearly independent correlation function in the evolution equations. In general, for the considered scalar field theory there are two linearly independent two-point functions, which may be associated to the anti-commutator of two fields ($F$) and the commutator expectation value: 
\begin{eqnarray}
&& \rho_{ab}(\tau,\tau',x_T- x_T',\eta-\eta') \nonumber\\ 
&& = i \left\langle\left[\hat{\phi}_a(\tau,x_T,\eta),\hat{\phi}_b(\tau',x_T',\eta')\right]\right\rangle \,.
\label{eq:defrho}
\end{eqnarray}
Here $\rho$ denotes the spectral function determined by the commutator $[.,.]$. Therefore, it encodes the equal-time field commutation relations, which read in Fourier-space with respect to the transverse spatial coordinates and rapidity  
\begin{eqnarray}
\rho_{ab}(\tau,\tau',p_T,\nu)|_{\tau=\tau'} & = & 0 \, , 
\nonumber\\ 
\partial_\tau\rho_{ab}(\tau,\tau',p_T,\nu)|_{\tau=\tau'}
& = & \frac{\delta_{ab}}{\tau}  \, ,
\label{eq:commrel}\\
\partial_\tau\partial_{\tau'}\rho_{ab}(\tau,\tau',p_T,\nu)|_{\tau=\tau'} & = & 0 \, .
\nonumber
\end{eqnarray}
The $\tau$-dependence of the commutator between the field and its conjugate momentum enters via the metric tensor, $\sqrt{-g} = \tau$~\cite{Tranberg,Stefan}. 
Since these relations are valid at all times, they also fix the initial conditions for the evolution of $\rho_{ab}(\tau,\tau',p_T,\nu)$. Again, using $O(N)$ symmetry we can write $\rho_{ab}=\text{diag}(\rho_{\|},\rho_{\bot},\ldots,\rho_{\bot})$.

Going beyond the linear regime using, e.g., the 2PI $1/N$-expansion to NLO the field $\phi$, the fluctuations $F_{\bot,\|}$ and the spectral functions $\rho_{\bot,\|}$ form a closed set of coupled evolution equations. The linear equations (\ref{eq:LinFT}) are generalized to their non-linear form, which reads for the longitudinal fluctuations
\begin{eqnarray}
&& \left[\partial_{\tau}^2+\frac{1}{\tau}\partial_{\tau}+p_T^2+\frac{\nu^2}{\tau^2} + M_{\|}^2 + \frac{\lambda}{2N}\phi^2(\tau)\right]F_{\|}(\tau,\tau^{\prime},p_T,\nu) 
\nonumber \\
&& = \, - \int_{\tau_0}^{\tau} d \tau^{\prime\prime} \tau^{\prime\prime}\, \Sigma^\rho_{\|}(\tau, \tau^{\prime\prime},p_T,\nu)\,
F_{\|}(\tau^{\prime\prime},\tau^{\prime},p_T,\nu) \nonumber\\
&& \quad + \, \int_{\tau_0}^{\tau^{\prime}} d \tau^{\prime\prime} \tau^{\prime\prime}\, \Sigma^F_{\|}(\tau, \tau^{\prime\prime},p_T,\nu)\,
\rho_{\|}(\tau^{\prime\prime},\tau^{\prime},p_T,\nu) \, .
\label{eq:nonLinFp} 
\end{eqnarray}
Here the effective mass term $M_{\|}^2 = M_{\|}^2(F_{\bot,\|})$ and the non-zero spectral and statistical parts of the self-energy 
$\Sigma^{\rho,F}_{\|}=\Sigma^{\rho,F}_{\|}(\rho_{\bot,\|},F_{\bot,\|},\phi)$ make the evolutions non-linear.\footnote{The spectral part, $\Sigma^\rho$, can be related to the imaginary part and the statistical part, $\Sigma^F$, to the real part of the self-energy for the considered theory~\cite{Berges:2004yj}.} The explicit linear $\tau^{\prime\prime}$-term in the integrand stems from the determinant of the metric tensor for the comoving coordinates. The spectral functions obey a similar equation with the characteristic 'memory integrals' over time, which reads for the longitudinal components
\begin{eqnarray}
&& \left[\partial_{\tau}^2+\frac{1}{\tau}\partial_{\tau}+p_T^2+\frac{\nu^2}{\tau^2} + M_{\|}^2 + \frac{\lambda}{2N}\phi^2(\tau)\right]\rho_{\|}(\tau,\tau^{\prime},p_T,\nu) 
\nonumber \\
&& = \, - \int_{\tau^\prime}^{\tau} d \tau^{\prime\prime} \tau^{\prime\prime}\, \Sigma^\rho_{\|}(\tau, \tau^{\prime\prime},p_T,\nu)\,
\rho_{\|}(\tau^{\prime\prime},\tau^{\prime},p_T,\nu)  \, .
\label{eq:nonLinrhop} 
\end{eqnarray}
The equivalent equations for the transverse components $F_\bot$ and $\rho_\bot$ can also be obtained from the corresponding linearized equations (\ref{eq:LinFT}) by replacing $m^2$ with an effective mass term $M_\bot^2(F_{\bot,\|})$ and taking into account a non-zero right hand side. The latter is of the same form as in (\ref{eq:nonLinFp}) and (\ref{eq:nonLinrhop}) with all longitudinal components replaced by transverse ones. At NLO in the 2PI $1/N$ expansion the remaining equation for the field $\phi$ can be written in the form
\begin{eqnarray}
&& \left[\partial_{\tau}^2+\frac{1}{\tau}\partial_{\tau}+M_{\|}^2+\frac{\lambda}{6N}\phi^2(\tau)\right]\phi(\tau) 
\nonumber \\
&& = \, - \int_{\tau_0}^{\tau} d \tau^{\prime} \tau^{\prime}\, \Sigma^\rho_{\|}(\tau, \tau^{\prime},p_T=0,\nu=0)|_{\phi=0}\,
\phi(\tau^\prime) \, ,
\label{eq:NLOphiEvoEq} 
\end{eqnarray}
where the spectral part of the self-energy is evaluated for zero field, i.e.\ $\Sigma^{\rho}_{\|}(\rho_{\bot,\|},F_{\bot,\|},\phi=0)$~\cite{Berges:2002cz,Berges:2004yj}. Of course, at even higher order in $1/N$ also additional terms that are non-linear in the field $\phi$ appear~\cite{JB:2PI1/N}.

\begin{figure}[t]
\includegraphics[scale=0.35]{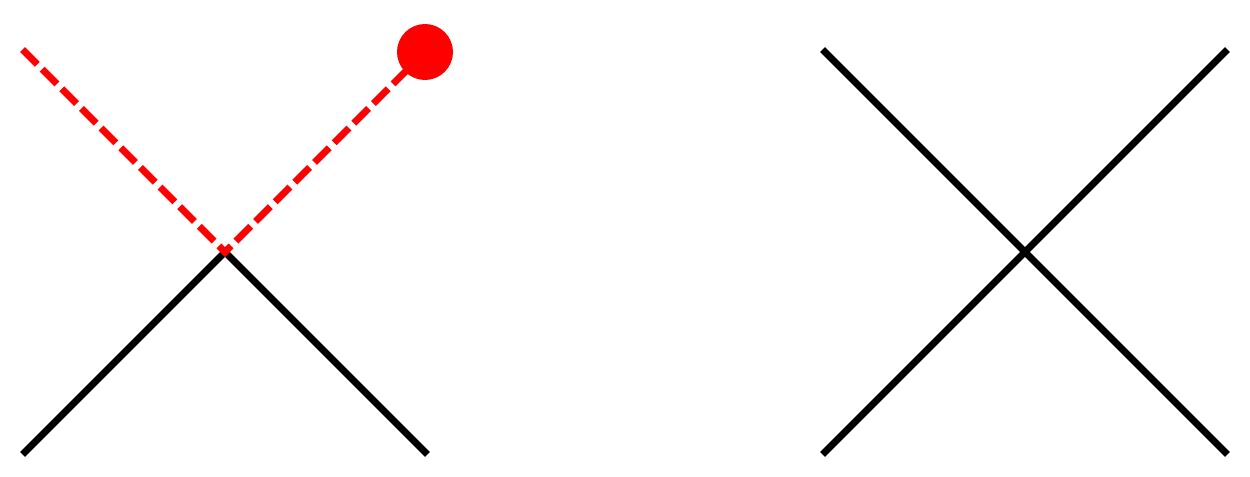}
\caption{\label{fig:vertices} (color online) Vertices in the presence of a macroscopic field. Dashed (red) lines denote 'longitudinal' field components and solid (black) lines are associated to either longitudinal or 'transverse' components. The dot indicates a non-zero field expectation value.} 
\end{figure}

We will start by classifying the nonlinear corrections to the evolution equations entering via the effective mass or self-energy terms $M_{\bot,\|}^2$, $\Sigma^{\rho}_{\bot,\|}$ and $\Sigma^{F}_{\bot,\|}$. In order to write down self-energies, it is important to note that in addition to the four-vertex proportional to $\lambda$ there is an effective three-vertex due to the presence of the macroscopic field. This is visualized in the left panel of Fig.~\ref{fig:vertices}, where dashed (red) lines denote 'longitudinal' ($a=1$) field components and solid (black) lines are associated to either longitudinal or 'transverse' ($a=2,\ldots,N$) components. A dot indicates a non-zero field expectation value $\phi(\tau)$ as in (\ref{eq:fielddirection}). 

A general non-linear contribution to the evolution equations will contain powers of $\lambda$, of the 'propagators' $F_{\bot,\|}$ and $\rho_{\bot,\|}$, and of the field $\phi$. Initially, the parametric dependence of the field is $\phi^2 \sim 1/\lambda$. Therefore, at not too late times a classification based on a small $\lambda$ has to take this into account. Most importantly, taking $F_{\bot,\|}$ into account for the power counting is crucial, since $F$ grows exponentially in time as a consequence of an instability. In contrast, the 'weight' of the spectral function $\rho_{\bot,\|}$ remains parametrically of order one at all times as is encoded in the equal-time commutation relations (\ref{eq:commrel})~\cite{Berges:2002cz}.
It is also important to note the fact that transverse fluctuations ($F_{\bot}$) exhibit the dominant growth in the linear regime. Consequently, contributions containing more transverse propagators ($F_{\bot}$) can become important earlier than those diagrams containing longitudinal propagators ($F_{\|}$) instead. For instance, an expression containing powers $\lambda^n F_{\bot}^m \phi^{2l}$ with integers $n,m$ and $l$ may be expected at not too late times to give sizable corrections to the linearized evolution equations once $F_{\bot} \sim 1/\lambda^{(n-l)/m}$ for typical momenta. Here $n$ yields the suppression factor from the coupling constant, whereas $m$ introduces the enhancement due to large fluctuations for typical momenta and $l$ due to a large macroscopic field. The power counting can become more involved as time proceeds, and it is remarkable that one can indeed identify a sequence of characteristic time scales with corresponding growth rates.

\subsection{Characteristic time scales and growth rates}

To start with a simple example, we consider first one-loop 'tadpole' corrections, which are obtained by 'closing' two longitudinal or two transverse 'legs' of the four-vertex in Fig.~\ref{fig:vertices} on the right. We emphasize already here that there are other corrections which will be of relevance before tadpoles come into play for not too large $N$. However, it will be convenient to express time scales in terms of the characteristic time when  tadpoles become relevant, since this turns out to coincide with the time when an infinite series of corrections become sizable and all exponential growth of fluctuations stops. Tadpole corrections are mass-like and their contribution reads 
\begin{eqnarray}
M^2_{\bot}(\tau) &=& m^2 + {\mathrm{T}}_{\|}(\tau) + (N+1) {\mathrm{T}}_{\bot}(\tau)
\, , \nonumber\\
M^2_{\|}(\tau) &=& m^2 + 3 {\mathrm{T}}_{\|}(\tau) + (N-1) {\mathrm{T}}_{\bot}(\tau)
\label{eq:effmass}
\end{eqnarray}
for the nonlinear evolution equation of the transverse ($F_{\bot}$) and the longitudinal fluctuations ($F_{\|}$), respectively. The one-loop tadpole integrals read
\begin{equation}
{\mathrm{T}}_{\bot,\|}(\tau) \, = \, \frac{\lambda}{6N}\int^{\Lambda}  \frac{d^2 p_T d\nu}{(2\pi)^3}\, F_{\bot,\|}(\tau,\tau,p_T,\nu) \, ,
\label{eq:tadpole} 
\end{equation}
where $\Lambda$ denotes some suitable regularization that enters the renormalization procedure, which has not to be specified for the current purpose. While initially the tadpole contributions (\ref{eq:tadpole}) are suppressed by the coupling constant $\lambda$, they become sizable at later times once the fluctuations grow parametrically to $F_{\bot,\|}(\tau,\tau,p_T,\nu) \sim 1/\lambda$ for typical momenta $p_T$ and $\nu$. More precisely, for the massless case ($m^2=0$) the size of a tadpole contribution in (\ref{eq:effmass}) should be compared to that of the macroscopic field-squared term $\lambda \phi^2(\tau)/6N$ in (\ref{eq:NLOphiEvoEq}), which on average is $\sigma_0^2/2~\left(\tau/\tau_0\right)^{-2/3}$ at early times. We will denote the time when both become of the same order of magnitude as $\tau_{\mathrm{nonpert}}$. Using that the tadpole will quickly be dominated by unstable modes entering the integral with characteristic primary growth-rate $\gamma(p_T,\nu,\theta)\leq\gamma_0$, such that
\begin{eqnarray}
\left|\frac{F(\theta,\theta',p_T,\nu)}{F(\theta_0,\theta_0,p_T,\nu)}\right|\leq\exp\left[\gamma_0\left(\theta+\theta'-2\theta_0\right)\right]\;,
\end{eqnarray}
 one obtains the estimate
\begin{eqnarray}
\label{eq:tnonpert}
\tau_{\mathrm{nonpert}} \, \gtrsim \, \tau_0\left[1+\frac{1}{3\sigma_0\tau_0\gamma_0}\text{ln}\left(\frac{\sigma_0^2}{2(N+1) {\mathrm{T}}_{\bot}(\tau_0)}\right)\right]^{3/2} \;. \nonumber \\
\end{eqnarray}
Here we used the important fact that transverse fluctuations exhibit the dominant growth in the linearized evolution equations. It is noteworthy that $\tau_{\mathrm{nonpert}}$ is rather sensitive to the inverse of the primary growth-rate, whereas the coupling constant and the size of the initial fluctuations only enter logarithmically through $T_{\bot}(\tau_0)$ according to (\ref{eq:tadpole}). In the weak coupling limit (\ref{eq:tnonpert}) reduces to
\begin{eqnarray}
\label{eq:tnonpertWC}
\tau_{\mathrm{nonpert}} \, \stackrel{(\lambda\ll1)}{\gtrsim}  \, \tau_0\left[1+\frac{1}{3\sigma_0\tau_0\gamma_0}\text{ln}\left(\frac{1}{\lambda}\right)\right]^{3/2} \;.
\end{eqnarray}
at leading logarithmic accuracy.

We strongly emphasize that when fluctuations have grown parametrically to ${\cal{O}}(1/\lambda)$ at $\tau_{\mathrm{nonpert}}$, it is not only a one-loop tadpole that becomes of order one. Since there is an infinite series of contributions including arbitrarily high loop-orders that become sizable, a non-perturbative approach such as the 2PI $1/N$ expansion or classical-statistical simulation methods have to be applied. Before addressing the non-perturbative regime in Sec.~\ref{sec:classicalstatistical} below, where no longer an exponential growth will be observed, we first consider earlier times than $\tau_{\mathrm{nonpert}}$ to discuss the nonlinear amplification of instabilities. More precisely, we will focus here on a subset of diagrams of the 2PI $1/N$ expansion to NLO \cite{JB:2PI1/N}, which are relevant for times $\tau \lesssim \tau_{\mathrm{nonpert}}$.

\begin{figure}[t]
\includegraphics[scale=0.25]{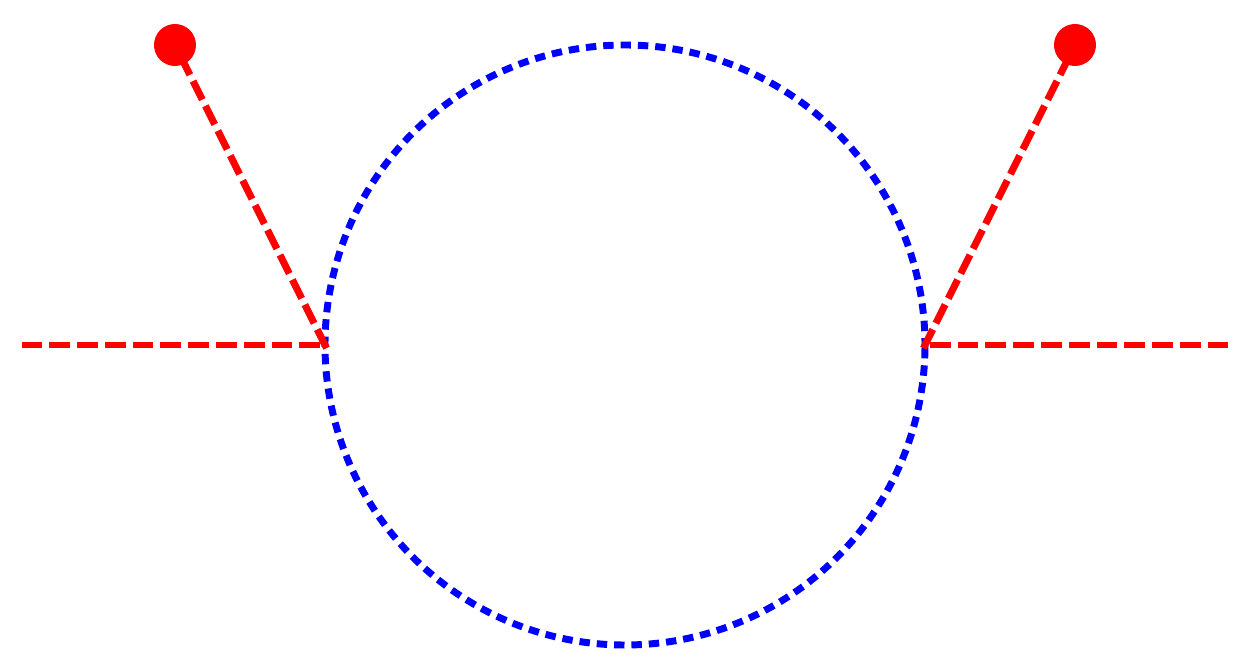}
\caption{\label{fig:NL1} (color online) One-loop contribution to the longitudinal component of the self-energy. Transverse two-point functions are denoted by dotted (blue) lines.}
\end{figure}
In general, the smaller the above introduced parameter $(n-l)/m$ for a specific self-energy correction, the earlier it may be expected to play a sizable role during the nonequilibrium time evolution. For the tadpoles this parameter is one. However, there is another one-loop self-energy correction for which this parameter is $1/2$, which solely contributes to the longitudinal components of the self-energy, as indicated by the red dashed (amputated) legs. The respective one-loop contribution to the longitudinal component of the self-energy is displayed graphically in Fig.~\ref{fig:NL1}. Here we denote transverse two-point functions by dotted (blue) lines. It is important to note that in this power counting scheme all other diagrams are suppressed by at least a fractional power of the coupling constant $\lambda$. Hence there exists a kinematic window where the only relevant self-energy correction originates from the diagram shown in Fig.~\ref{fig:NL1}. In the following, we will see that this leads already to a non-linear amplification of the primary instability for longitudinal fluctuations. In particular, this correction has to be taken into account when considering further corrections that become relevant at later times. 

The spectral and statistical self-energies associated to Fig.~\ref{fig:NL1} read
\begin{eqnarray}
&& \Sigma^{\rho}_{\|}(\tau,\tau^{\prime\prime},p_T,\nu) \, \stackrel{\text{(one-loop)}}{=} \, -\frac{4\lambda(N-1)}{6N}\sigma(\tau)\sigma(\tau^{\prime\prime}) \nonumber\\ && \times \int\frac{d^2 q_T d\nu_q}{(2\pi)^3}\,\rho_{\bot}(\tau,\tau^{\prime\prime},p_T-q_T,\nu-\nu_q)F_{\bot}(\tau,\tau^{\prime\prime},q_T,\nu_q) \, ,
\nonumber\\
&& \Sigma^{F}_{\|}(\tau,\tau^{\prime\prime},p_T,\nu) \, \stackrel{\text{(one-loop)}}{=} \, -\frac{2\lambda(N-1)}{6N}\sigma(\tau)\sigma(\tau^{\prime\prime})  \nonumber\\
&& \times \int\frac{d^2 q_T d\nu_q}{(2\pi)^3} \Big[ F_{\bot}(\tau,\tau^{\prime\prime},p_T-q_T,\nu-\nu_q)F_{\bot}(\tau,\tau^{\prime\prime},q_T,\nu_q) \nonumber \\
&& -\frac{1}{4} \rho_{\bot}(\tau,\tau^{\prime\prime},p_T-q_T,\nu-\nu_q)\rho_{\bot}(\tau,\tau^{\prime\prime},q_T,\nu_q)\Big]. 
\label{eq:NL1:SEF}
\end{eqnarray}
These enter the memory integrals on the right-hand-side of the evolution-equations (\ref{eq:nonLinFp}), (\ref{eq:nonLinrhop}) and (\ref{eq:NLOphiEvoEq}). We note that the ($\rho_\bot \rho_\bot$)-term in the integrand for $\Sigma^{F}_{\|}$ is a genuine quantum correction, which would be absent in a classical-statistical description~\cite{Berges:2004yj}. However, since $F_\bot F_\bot \gg \rho_\bot \rho_\bot$ once nonlinear corrections become sizable, one can neglect the quantum part to very good accuracy. We will exploit this fact further in Sec.~\ref{sec:classicalstatistical} on classical-statistical simulations below.

In order to make analytical progress, one can exploit the fact that the dominant contribution to the memory integrals originates from late times when fluctuations have become exponentially large~\cite{Berges:2002cz}. Instead of considering integrals from $\tau_0$ to $\tau$ and $\tau'$, respectively, we consider the integrals only over some suitable, small interval $\Delta$. This will be sufficient to obtain characteristic time scales to leading logarithmic accuracy.
It allows us to expand the integrand around the times of interest, where at leading order one finds
\begin{eqnarray}
F(\tau,\tau'',p_T,\nu) &\simeq& F(\tau,\tau,p_T,\nu) \, ,\\
\rho(\tau,\tau'',p_T,\nu) &\simeq& \frac{\tau-\tau''}{\tau} \simeq \frac{\tau-\tau''}{\tau''} \, .
\end{eqnarray}
Here we used the equal-time commutation relations (\ref{eq:commrel}) to expand the spectral function. With these approximations one can explicitly evaluate the right-hand-side of the evolution equation (\ref{eq:nonLinFp}) as
\begin{eqnarray}
&& {\text{RHS}} \, \simeq \, \Delta^2\, \frac{\lambda(N-1)}{3N}\, \sigma^2(\tau) \int\frac{d^2 q_T d\nu_q}{(2\pi)^3} F_{\bot}(\tau,\tau,q_T,\nu_q) 
\nonumber\\
&& \times\,\,  F_{\|}(\tau,\tau',p_T,\nu) + \Delta^2\, \frac{\lambda(N-1)}{6N}\, \sigma(\tau)\sigma(\tau')
\label{eq:NL1:source}\\
&& \times\, \int\frac{d^2 q_T d\nu_q}{(2\pi)^3} F_{\bot}(\tau,\tau',p_T-q_T,\nu-\nu_q)F_{\bot}(\tau,\tau',q_T,\nu_q)\,. 
\nonumber\end{eqnarray}
The first term in the above sum acts as a momentum-independent mass term similar to the tadpole term on the LHS of the evolution equation (\ref{eq:nonLinFp}). For it to be relevant it requires $F_{\bot} \sim \mathcal{O}(1/\lambda)$ as discussed before and, hence, one expects a relevant contribution staring at $\tau_{\mathrm{nonpert}}$. In contrast, the second term appearing in (\ref{eq:NL1:source}) acts as a source term for the evolution of the longitudinal fluctuations $F_{\|}$. The momentum dependence is given by the convolution of the transverse fluctuations $F_{\bot}$ with itself. For instance, if $F_{\bot}(\tau,\tau,q_T,\nu_q)$ is peaked around some momenta $p_{0}$ and $\pm\nu_0$, the source term has its dominant contributions around $p_T=\{0,2p_{0}\}$ and $\nu=\{0,\pm2\nu_0\}$. 
\begin{figure}[t]
\includegraphics[scale=0.5]{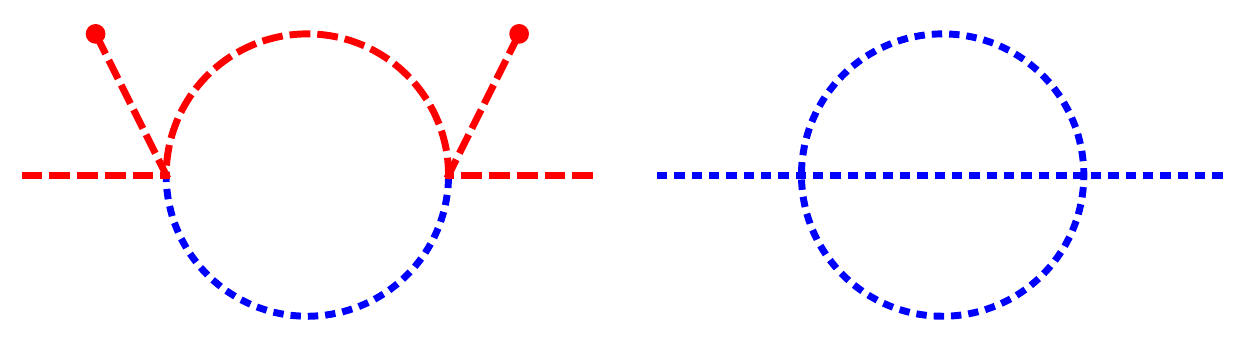}
\caption{\label{fig:NL2} (color online) One- and two-loop contributions to the transverse components of the self-energies.}
\end{figure}

In the presence of this strong source term, one expects $F_{\|}$ to follow the source $\sim \lambda F_{\bot}^2$. In this way the primary growth of the transverse fluctuations $F_{\bot}$ leads after some delay-time $\tau_{\mathrm{source}}$ to a secondary stage of growth, where the longitudinal modes $F_{\|}$ exhibit amplified growth-rates. In particular for the longitudinal zero-mode, i.e.\ $F_{\|}(\tau,\tau',p_T=0,\nu=0)$, the secondary growth-rate can be twice as large as the primary growth-rate observed for $F_{\bot}$. It is important to note here that the amplification occurs before tadpole corrections become important. This is due to the fact that the source term is of order $\lambda F_{\bot}^2$ which requires $F_{\bot} \sim \mathcal{O}(1/\lambda^{1/2})$ for typical momenta to yield relevant contributions. With the same approximations entering the estimate for $\tau_{\mathrm{nonpert}}$ one can estimate the associated time-scale as
\begin{eqnarray}
\label{eq:tsource}
\tau_{\mathrm{source}}\gtrsim\tau_0\left[1+\frac{1}{6(\sigma_0\tau_0)\gamma_0}\text{ln}\left(\frac{1}{2(N-1) T(\tau_0)\Delta^2}\right)\right]^{3/2} \;. \nonumber \\
\end{eqnarray}
Again, in the weak coupling limit the dominant contribution arises from $\text{ln}(\lambda^{-1})$ and (\ref{eq:tsource}) reduces to
\begin{eqnarray}
\tau_{\mathrm{source}}\stackrel{(\lambda \ll 1)}{\simeq}\frac{\tau_{\mathrm{nonpert}}}{2^{3/2}} \, , 
\end{eqnarray}
which is smaller than the time-scale on which screening effects due to effective mass terms become relevant. Hence, there is a period of time when one expects growth of the longitudinal modes due to a nonlinear amplification of the primary instability. This happens for a bound momentum region which is entirely determined by the spectral shape of the primary instability. We emphasize that the larger couplings and/or initial fluctuations the earlier this non-linear amplification of the instability happens.

We have seen that as a consequence of nonlinear amplifications characteristic longitudinal fluctuations can be expected to become $\mathcal{O}(\lambda F^2_{\bot})$ around the time $\tau_{\mathrm{source}}$. This modifies the power counting for subsequent times, since in addition to parametrically large $F_{\bot}$ also parametrically large $F_{\|}$ enter loop corrections. In particular, this will soon after lead to sizable nonlinear contributions to the evolution equation for transverse fluctuations $F_\bot$. The relevant diagrams contributing to the transverse components of the self-energies are displayed in Fig.~\ref{fig:NL2}. Assuming $F_{\|}\sim\mathcal{O}(\lambda F^2_{\bot})$ for typical momenta, both depicted diagrams are $\mathcal{O}(\lambda^2 F_{\bot}^3)$. Thus one expects a relevant contribution as soon as $F_{\bot}\sim1/\lambda^{2/3}$. The momentum-dependence of the diagrammatic contribution leads to an extension of the amplified region to higher momenta, where again multiples of the primary growth-rate appear. It is important to realize that this amplification repeats itself, i.e.\ the newly amplified modes together with the primarily amplified ones act as a source for other modes. In this way the instability propagates to higher and higher momentum modes for both longitudinal and transverse fluctuations. This can be nicely observed from our numerical simulations, which are presented in the next section. 

\section{Classical-statistical lattice simulations}
\label{sec:classicalstatistical}

\subsection{Comparison with analytic results}

As the system evolves in time, occupation numbers may grow until the fluctuations $F_{\bot,\|}$ become as large as $\mathcal{O}(1/\lambda)$ around the time $\tau_{\mathrm{nonpert}}$ given by (\ref{eq:tnonpert}). Hence, there are sizable contributions to the dynamics originating from all loop orders and a non-perturbative description is necessary. The growth saturates at this time and a comparatively slow, quasistationary evolution sets in. This has been discussed in great detail for non-expanding systems \cite{Berges:2002cz}. In that context it has also been shown that classical-statistical approximations can reproduce very well the results from the non-perturbative $1/N$ expansion to NLO of the quantum 2PI effective action at not too late time, i.e.\ before the approach to quantum thermal equilibrium sets in~\cite{Aarts:2001yn,TranbergSmit,Berges:2008wm}.

\begin{figure}
\includegraphics[scale=0.7]{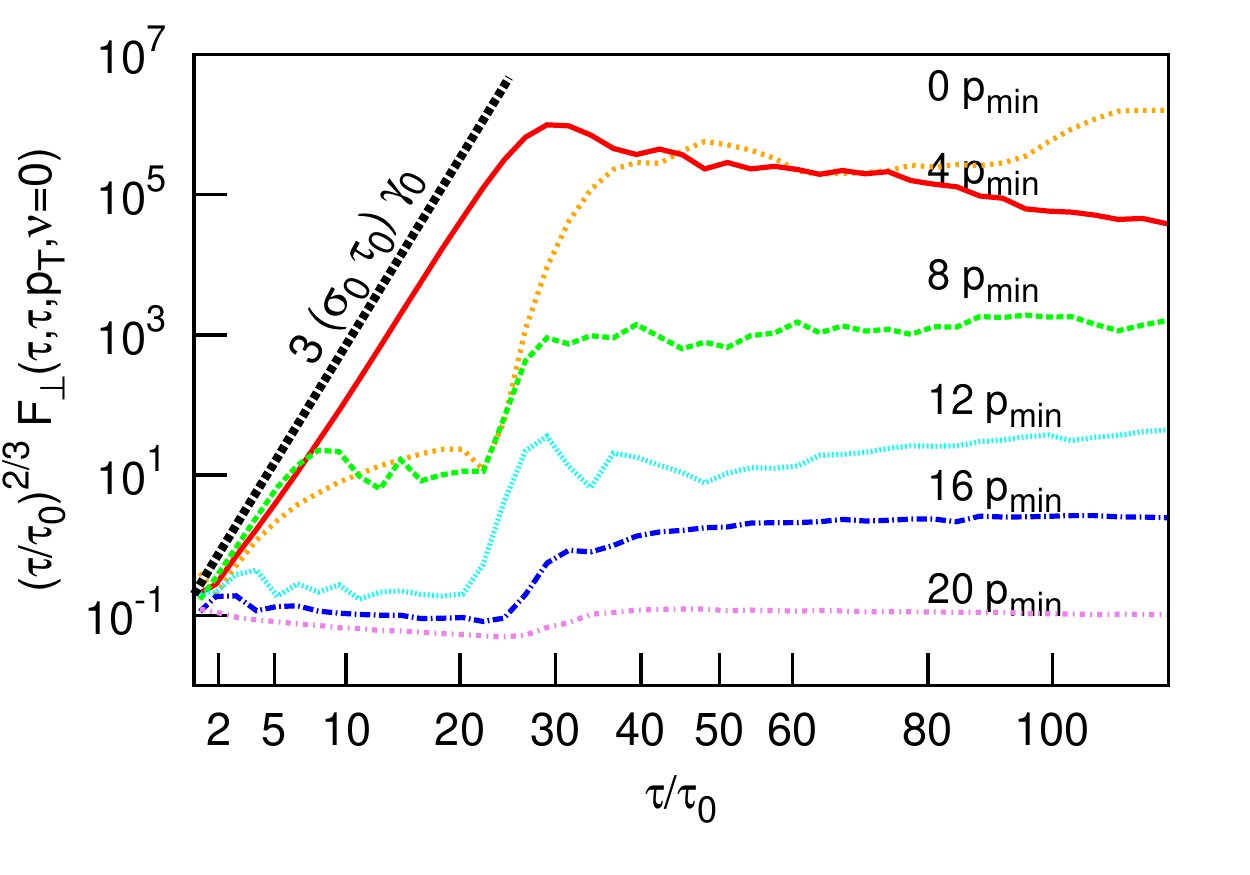}
\caption{(color online) \label{fig:FTpt} Time evolution of transverse fluctuations with $p_T=0,4,8,12,16,20~p_{\text{min}}$ and $\nu=0$ averaged over one period of oscillation of the macroscopic field.}
\end{figure}

In this section we present the results obtained from classical-statistical lattice simulations for the longitudinally expanding case discussed above. These are described by numerically solving the classical field equations of motion, obtained from the stationarity of the action (\ref{eq:action}), and Monte Carlo sampling of initial conditions such that the initial classical averages agree with the quantum initial correlation functions. The parameters of these simulations are chosen as $m^2=0$, $\sigma_0\tau_0=5$, $N=4$ and $\lambda=10^{-4}$. We discretize the evolution equation for the inhomogeneous classical field on a three-dimensional lattice in transverse coordinates and rapidity with grid size $N_t^2 \times N_{\eta}$. The lattice spacing is chosen such that the lattice ultraviolet cut-offs $\sim 1/a_t$ and $\sim 1/(\tau a_{\eta})$ are above all physical scales. Of course, for the expanding system this condition is time-dependent through the time-dependence of the lattice cut-off due to red-shift as well as the time-dependence of physical scales due to dilution. Numerical simulations at late times are, therefore, computationally hard to perform and we will focus the discussion on the physics at sufficiently early times. If not stated otherwise we will use $N_t=N_{\eta}=128$ with the discretization $\sigma_0 a_t=0.5$ and $a_\eta=0.1$. For convenience we express the transverse and longitudinal momenta in terms of the lattice momenta $p_{min}=2\sqrt{2}/(N_t a_t)$ and $\nu_{min}=2.0/(N_{\eta} a_{\eta})$, respectively.

\begin{figure}[t]
\includegraphics[scale=0.7]{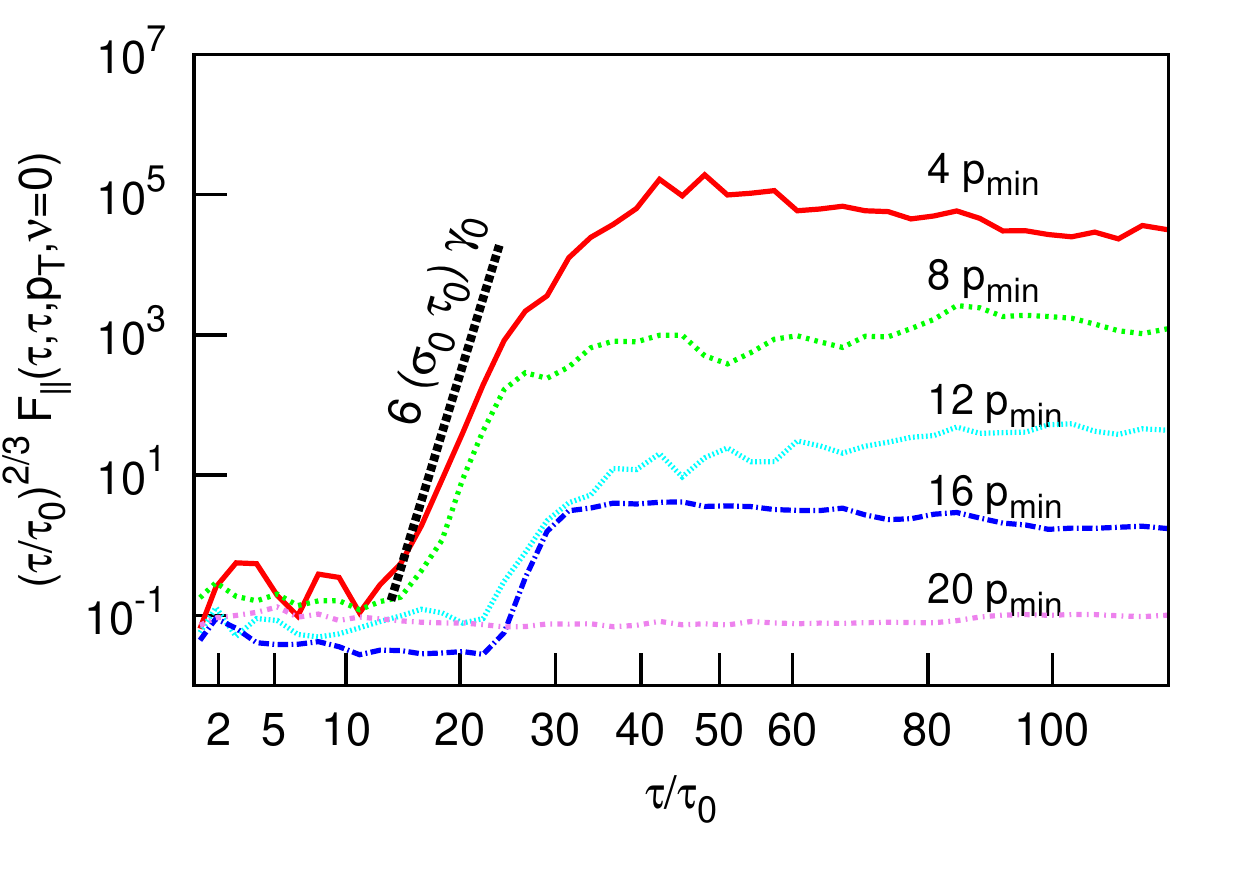}
\caption{ (color online) \label{fig:FLpt} Time evolution of longitudinal fluctuations with $p_T=4,8,12,16,20~p_{\text{min}}$ and $\nu=0$ averaged over one period of oscillation of the macroscopic field.}
\end{figure}

\begin{figure}[t]
\includegraphics[scale=0.7]{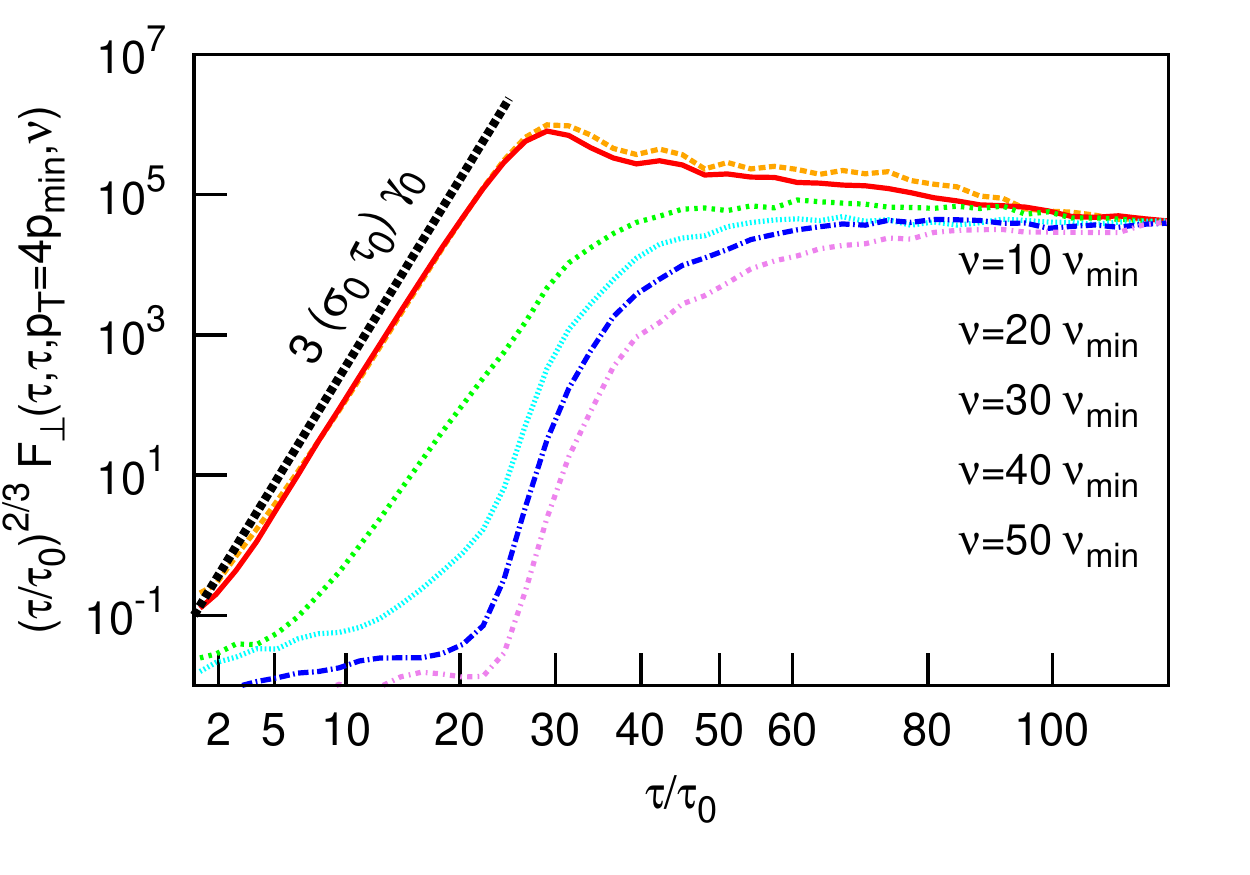}
\caption{ (color online) \label{fig:FTnu} Time evolution of transverse fluctuations with $p_T=4~p_{\text{min}}$ and $\nu=0,10,20,30,40,50~\nu_{\text{min}}$ (top to bottom) averaged over one period of oscillation of the macroscopic field.}
\end{figure}

The time-evolution of transverse and longitudinal fluctuations for different momenta is presented in Figs.~\ref{fig:FTpt} and \ref{fig:FLpt} for different transverse momenta $p_T$ and rapidity wave number $\nu=0$. Fig.~\ref{fig:FTnu} shows the transverse fluctuations for modes with different rapidity wave numbers $\nu$ and transverse momentum $p_T=4~p_{\text{min}}$ for which the leading primary instability occurs. From Fig.~\ref{fig:FTpt} one observes that the primary instability occurs for a bound momentum region. The phenomenon of linear freeze-out is clearly visible. The modes with $p_T=8~p_{\text{min}}~\text{and}~12~p_{\text{min}}$, for instance, exhibit exponential growth at early times but decouple from the instability shortly. The functional form of the primary instability is well described by an exponential in $(\tau/\tau_0)^{2/3}$ with maximum growth-rate $3 \sigma_0\tau_0 \gamma_0$.

At later times, when the primary instability has been operative long enough to produce large transverse fluctuations, secondary instabilities set in for the longitudinal modes, as can be observed from 
Fig.~\ref{fig:FLpt}. These secondary instabilities exhibit growth rates up to $6 \sigma_0\tau_0 \gamma_0$ as discussed in Sec.~\ref{sec:nonlinear}. The onset of these secondary instabilities is limited to a momentum region $p_T\lesssim 8~p_{\text{min}}$ as they originate from the two-loop diagram shown in Fig.~\ref{fig:NL2}. This first non-linear amplification of the primary instability happens at times $\tau/\tau_0\approx15$. At later times $\tau/\tau_0\approx20$ the system exhibits collective amplification of the primary instability and fluctuations begin to grow in a wide momentum range. Also modes that previously exhibited freeze-out show a second period of growth while modes with smaller primary growth-rates exhibit a significant speed-up. One can also observe from Figs.~\ref{fig:FTpt} and \ref{fig:FLpt} how this sets off an avalanche of instabilities propagating to higher momenta. At this point of the evolution, modes with high rapidity wave number $\nu$ set in earlier as suggested by the linearized evolution equations. From Fig.~\ref{fig:FTnu} one observes that once amplification by nonlinear corrections sets in, modes in a large momentum region show significant growth within a very short time. This happens because high longitudinal momenta no longer have to wait to become unstable, but are instead subject to the nonlinear amplification. This is indeed very similar to what is observed in numerical simulations of pure gauge theory with 'Glasma' initial conditions \cite{Romatschke:2005pm}. Ultimately the growth of the instability saturates when the fluctuations become of the order of the inverse coupling $\mathcal{O}(1/\lambda)$. The rapid dynamics of instabilities is then followed by a regime of comparatively slow and smooth evolution, where one expects transport approaches to be applicable.

\subsection{Energy transfer to particles and EOS}

\begin{figure}[tp]
\begin{minipage}[b]{0.5\textwidth}
\centering
\includegraphics[width=\textwidth]{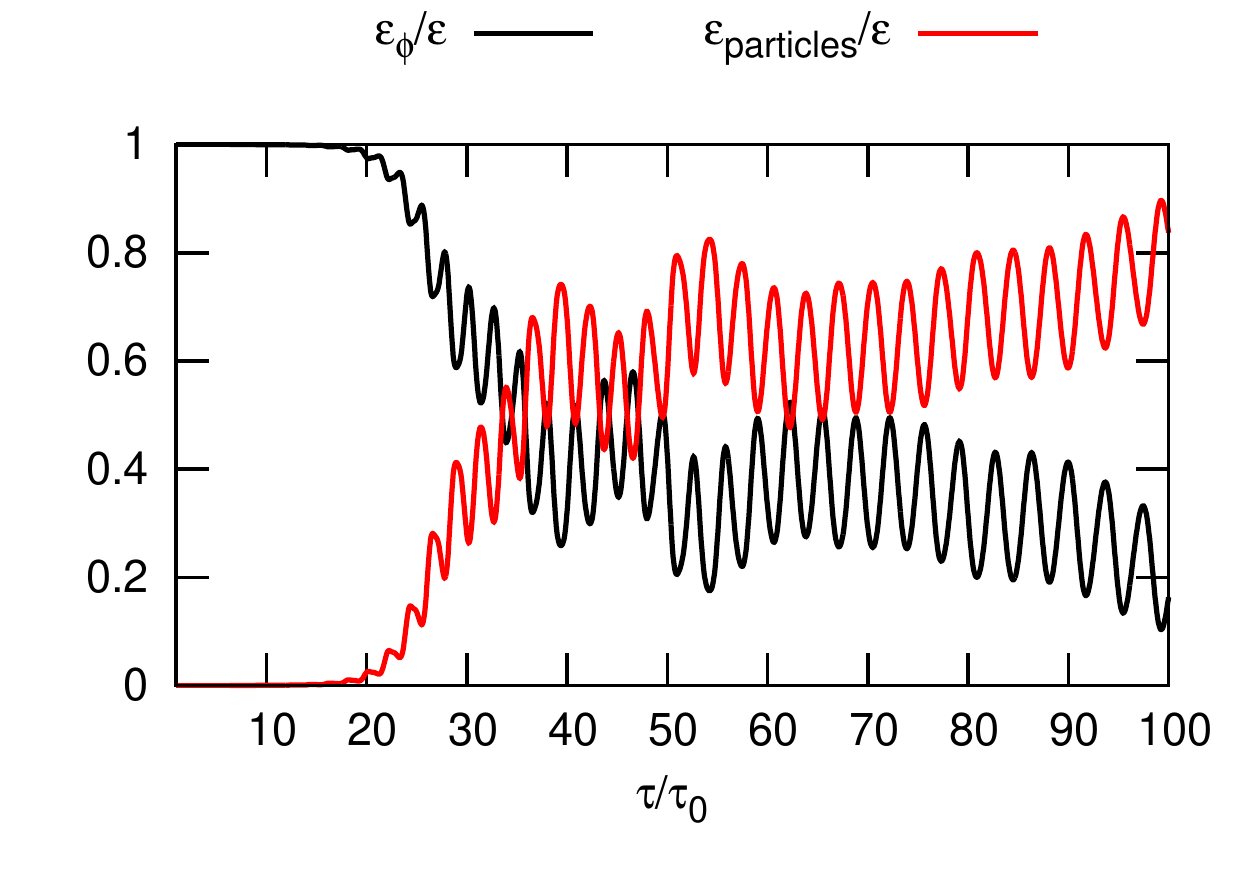}
\end{minipage}
\begin{minipage}[b]{0.5\textwidth}
\centering
\includegraphics[width=\textwidth]{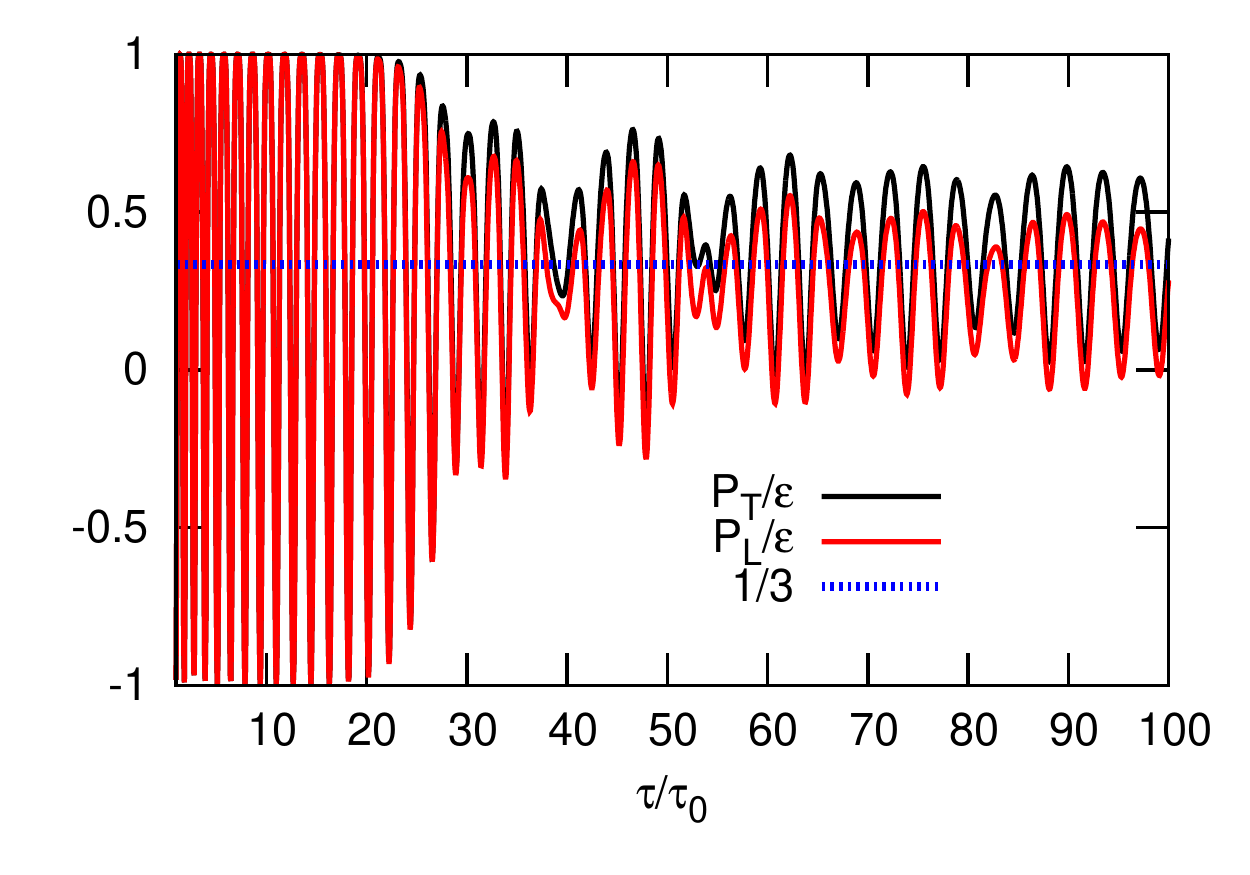}
\end{minipage}
\caption{(color online) \label{fig:Energy}(\textbf{top}) Fraction of total energy contained in the macroscopic field and in the produced particles. Significant energy transfer to the produced particles occurs when high momentum modes carrying more energy are affected by secondary instabilities due to nonlinear amplification in a broad momentum range.(\textbf{bottom}) Dynamical equation of state for transverse and longitudinal pressure.
}
\end{figure}

Finally one is interested in the applicability of hydrodynamics to provide a simple description of the system under consideration. A necessary condition for this to be the case is the establishment of an equation of state relating pressure and energy density. To investigate this in our numerical simulations we consider the different components of the stress energy tensor
\begin{eqnarray}
T^{\mu\nu}(x)=\langle \left(\partial^{\mu}\varphi_a(x)\right)\left(\partial^{\nu}\varphi_a(x)\right)-g^{\mu\nu}\mathcal{L}[\varphi](x)\rangle \;.
\label{eq:Tmunu}
\end{eqnarray}
The quantities of interest here are the energy density $\epsilon$ as well as the longitudinal and transverse pressures, $P_L$ and $P_T$, defined by.\footnote{Since we assumed a boost-invariant system the local restframe is always given by $u^{\mu}(x)=(\cosh(\eta),0,0,\sinh(\eta))$ Therefore, one obtains energy-density and pressure immediately in terms of the components of the stress-energy tensor in comoving coordinates.}
\begin{eqnarray}
\epsilon(\tau)&=&\frac{1}{V}\int d^2 x_T d\eta\, T^{\tau\tau}(\tau,x_T,\eta) \, , \nonumber \\
P_T(\tau)&=&\frac{1}{V}\int d^2 x_T d\eta\, \frac{T^{11}(\tau,x_T,\eta)+T^{22}(\tau,x_T,\eta)}{2} \, ,\nonumber \\
P_L(\tau)&=&\frac{\tau^2}{V}\int d^2 x_T d\eta\, T^{\eta\eta}(\tau,x_T,\eta) \, .
\end{eqnarray}
To further investigate the energy transfer from the macroscopic field to the fluctuations, we define the energy content of the macroscopic field by replacing the fluctuating field in (\ref{eq:Tmunu}) by its expectation value, i.e.
\begin{eqnarray}
T^{\mu\nu}_{\phi}(x)= \left(\partial^{\mu}\phi_a(x)\right)\left(\partial^{\nu}\phi_a(x)\right)-g^{\mu\nu}\mathcal{L}[\phi](x) \; .
\end{eqnarray}
The energy contained in the fluctuations is then simply the difference between the two, i.e.
\begin{eqnarray}
T^{\mu\nu}_{\text{particles}}(x)=T^{\mu\nu}(x)-T^{\mu\nu}_{\phi}(x) \;.
\end{eqnarray}
In the upper part of Fig.~\ref{fig:Energy} we show the energy fractions contained in the macroscopic field and the produced particles as a function of time. One observes that the energy contained in the produced particles becomes significant around $\tau/\tau_0\approx20$, which coincides with the onset of nonlinear amplification in a broad momentum range discussed above. While the large population of soft particles is created at early times through the primary instability as suggested by Figs.~\ref{fig:FTpt} and \ref{fig:FTnu}, the energy transfer to the particles is to be understood as a consequences of hard particles being affected by secondary instabilities through collective amplification because of the enhanced phase-space and larger mode energy for hard particles \footnote{In a pertrubative treatment one has $T^{\mu\nu}(X)=\int \frac{d^4p}{(2\pi)^4} p^{\mu}p^{\nu} F(X,p)$, where $F(X,p)$ is the Wigner transform of the statistical two point correlation function \cite{Berges:2004yj}.}.

The ratios of longitudinal and transverse pressure over energy density, i.e.\ the dynamical equation of state, are presented in the lower panel of Fig.~\ref{fig:Energy} as a function of proper-time. Initially this observable is dominated by the energy contained in the macroscopic field. Therefore one observes large oscillations around the average value of 1/3. At later times when energy is carried predominantly by the produced particles the oscillations are damped and one observes a smooth behavior. In our simulations we find that the system exhibits a remaining anisotropy as longitudinal and transverse pressure do not equilibrate to 1/3. It is important to realize here that isotropy can only persist when the evolution of the system is dominated by particle interactions in contrast to the free-streaming behavior. The latter drives the system away from isotropy due to the red-shift of longitudinal modes. These quantitative aspects depend, of course, on the specific model and can be very different in the relevant case of QCD. At late times one should also employ the quantum evolution based on the 2PI $1/N$ expansion to NLO instead of classical-statistical simulations, since deviations can be expected for a dilute system. Unfortunately, there is no corresponding nonperturbative expansion so far that is manageable in QCD and this topic is beyond the scope of the present work.

\section{Conclusion and Outlook}
\label{sec:conclusion}

We have shown for the example of $N$-component scalar field theories that nonequilibrium instabilities can lead to important nonlinear phenomena due to quantum corrections for longitudinally expanding systems. At early times fluctuations are amplified due to primary instabilities, which are described by the linearized evolution equations. When these become large enough to induce nonlinear corrections, the instability propagates towards higher momentum modes and secondary growth-rates much higher than the primary ones can be observed. This way significant energy densities carried by the produced particles can be achieved, which has important consequences for the question of isotropization and the establishment of an equation of state.

The presence of secondary instabilities is rather insensitive to the detailed mechanism for primary growth of fluctuations. Qualitatively, very similar phenomena can be expected for nonabelian gauge theories as confirmed already for the non-expanding case \cite{Berges:2007re,Fujii:2009kb}. This has in principle been observed also in classical-statistical simulations of the Glasma \cite{Romatschke:2005pm,Fukushima:2011nq} and corresponding 2PI effective action techniques \cite{Hatta:2PICGC} may provide a systematic framework to study these effects. Even for the physics of primary instabilities with anisotropic Bjorken expansion many qualitative aspects, such as the delayed set-in of high $\nu$ modes observed in classical-statistical lattice simulations in SU(2) Yang-Mills theory \cite{Romatschke:2005pm}, can also be found in the scalar theory and explained analytically within our model.

The relevant time-scale for nonlinear effects is governed by the growth rates of the primary instability and to a weaker extent by the logarithmic dependence on the coupling constant and the spectrum of initial fluctuations. The evolution of the Glasma in the context of relativistic heavy ion-collisions may be strongly modified once realistic initial fluctuations are taken into account~\cite{CGC:IC}.
Here one expects the instability, which affects initially only small longitudinal momentum modes, to extend quickly to higher momenta by the nonlinear amplification mechanism if their nonzero initial amplitude is properly taken into account. 

In this work, we have not discussed the possible approach to a turbulent scaling regime once nonlinear corrections become of order one. It has been argued in the context of QCD that also a Bose condensate may form later on \cite{Blaizot:2011xf}. For the non-expanding case it has recently been shown for scalar theories that indeed a Bose condensate occurs as a consequence of an inverse particle cascade with a universal power-law spectrum \cite{Berges:2012us,Gelis:PR1}. This particle transport towards low momenta is part of a dual cascade, in which energy is also transfered by weak wave turbulence towards higher momenta. Similar phenomena have also been studied for cold atoms~\cite{Scheppach:2009wu,Berges:2012us}. Classical-statistical simulations including the relevant physics of longitudinal expansion~\cite{Romatschke:2005pm,Fukushima:2011nq} should be the appropriate theoretical approach to clarify these questions in the context of heavy-ion collisions at sufficiently high energies.\\

This work is supported in part by the BMBF grant 06DA9018 and by EMMI.

\section*{Appendix A: Conformal time in anisotropically expanding geometry}

Here we introduce the conformal time on the level of the classical action. This has the advantage that the scale factor appearing in the 2PI evolution equations (\ref{eq:nonLinFp})-(\ref{eq:NLOphiEvoEq}) can be absorbed into a redefinition of the time-variable and a rescaling of field variables and propagators. In contrast to the main text we do not make the conformal time variable dimensionless. The classical action in comoving coordinates is given by (\ref{eq:action}) 
\begin{eqnarray}
S[\varphi]&=&\int~d^2x_T~d\eta~d\tau~a(\tau) \left(\frac{1}{2}g^{\mu\nu}(x)\left(\partial_{\mu}\varphi_{a}\right)\left(\partial_{\nu}\varphi_{a}\right)\right. \nonumber \\
&&-\left.\frac{1}{2}m^{2}\varphi_{a}\varphi_{a}-\frac{\lambda}{4!N}\left(\varphi_{a}\varphi_{a}\right)^{2}\right)
\end{eqnarray}
with the metric $g^{\mu\nu}(x)=\text{diag}(1,-1,-1,-\tau^2)$ and the scale factor $a(\tau)$ which in the case of a one-dimensional Bjorken expansion is given by the determinant of the metric $a(\tau)=\sqrt{-g_{\mu\nu}(x)}=\tau$. In isotropically expanding backgrounds it is a frequently used concept to introduce conformal time variables to solve the dynamics in a quasi-static framework. Here we employ an approach that is very similar in spirit. By introducing a new time variable and rescaling the field variables according to 
\begin{equation}
d\tau=a^{1/3}\left(\tau\right)d\theta \;, \qquad \tilde{\varphi}_{a}=a^{1/3}(\tau)\varphi_{a} \;,
\end{equation}
where $\theta$ will be referred to as conformal time, the action can be rewritten as
\begin{eqnarray}
S[\tilde{\varphi}]&=&\int~d^2x_T~d\eta~d\theta \left(\frac{1}{2}\tilde{g}^{\mu\nu}(x)\left(\partial_{\mu}\tilde{\varphi}_{a}\right)\left(\partial_{\nu}\tilde{\varphi}_{a}\right)\right. \nonumber \\
&&-\left.\frac{1}{2}\tilde{m}^{2}(\theta)\tilde{\varphi}_{a}\tilde{\varphi}_{a}-\frac{\lambda}{4!N}\left(\tilde{\varphi}_{a}\tilde{\varphi}_{a}\right)^{2}\right)\;,
\label{eq:confAction}
\end{eqnarray}
where the determinant of the metric drops out and first order derivatives are absorbed into a redefinition of the effective mass through integration by parts. Here the metric is given by $\tilde{g}_{\mu\nu}=\text{diag}(1,-a^{-2/3}(\tau),-a^{-2/3}(\tau),-a^{4/3}(\tau))$. The mass term appearing in (\ref{eq:confAction}) depends explicitly on conformal time and is given by
\begin{equation}
\tilde{m}^{2}\left(\theta\right)=m^{2}a^{2/3}(\theta)+\frac{2}{9}\left(\frac{a'(\theta)}{a(\theta)}\right)^{2}-\frac{1}{3}\frac{a''(\theta)}{a(\theta)}\;,
\end{equation}
where primes denote derivatives with respect to conformal time. The last two terms originate from interchanging time-derivatives with the scale factor
\begin{eqnarray}
\int~d\tau~a(\tau)\left(\partial_{\tau}\varphi_{a}\right)^{2}
=\int d\theta\left(\tilde{\varphi}_{a}'-\frac{1}{3}\frac{a'(\theta)}{a(\theta)}\tilde{\varphi}_{a}\right)^{2}\;,
\end{eqnarray}
and integrating first derivatives by parts according to 
\begin{equation}
\int~d\theta~\frac{a'(\theta)}{a(\theta)}\varphi_{a}\varphi_{a}'=-\frac{1}{2}\int~d\theta\left(\frac{a''(\theta)}{a(\theta)}-\left(\frac{a'(\theta)}{a(\theta)}\right)^{2}\right)\varphi_{a}^{2}\;.
\end{equation}
The scale factor and its derivatives with respect to conformal time are explicitly given by
\begin{eqnarray}
a(\theta)&=&\left(\frac{2}{3}\theta\right)^{3/2} \;,
a'(\theta)=\left(\frac{2}{3}\theta\right)^{1/2} \;,
 \nonumber \\
a''(\theta)&=&\frac{1}{3}\left(\frac{2}{3}\theta\right)^{-1/2}\;.
\end{eqnarray}
The major advantage of the formulation in terms of conformal variables is the fact that the metric determinant drops out of the action. Therefore the evolution equations derived from the action (\ref{eq:confAction}) do not contain first order time derivatives. However the mass term $m^2(\theta)$ depends explicitly on time and dominates on large time-scales, indicating the freeze-out, unless $m^2=0$. In contrast for the massless case one finds that $m^2(\theta)$ decreases as a power of $\theta$ on large time-scales. Hence the conformal time formulation is particularly well suited for this situation, which is relevant also for gauge theories.

\section*{Appendix B: Solution of the Lam\'{e} equation}

Here we give some details for the solution of the Lam\'{e} equation as it appears in parametric resonance in Minkowski space-time needed for the main text. The discussion is presented for the massless case $m^2=0$ and closely follows in part Ref.~\cite{Boyanovski}. In this situation the linearized evolution equation for the fluctuations for transverse modes is given by
\begin{eqnarray}
\left[\partial_{\theta}^2 + p^2 +
\text{cn}^2 \left(\theta~;\frac{1}{2}\right)\right]F(\theta,\theta',p)=0.
\label{eq:LameCN}
\end{eqnarray}
By introducing the mode functions $f_p(\theta)$ according to
\begin{eqnarray}
F(\theta,\theta',p)=\frac{1}{2}\left[f_p(\theta)f_p^{*}(\theta')+f_p^{*}(\theta)f_p(\theta')\right] 
\end{eqnarray}
the partial differential equation (\ref{eq:LameCN}) can be transformed into a set of two independent ordinary differential equations for evolution in $\theta$ and $\theta'$ respectively. By use of the identities \cite{AbramovichStegun}
\begin{eqnarray}
\text{cn}^2(\theta,\alpha)&=&1-\text{sn}^2(\theta,\alpha) \, , \label{eq:Jprop1}\\
\text{sn}^2(\theta,\alpha)&=&\frac{1}{\alpha~\text{sn}^2(\theta+iK'(\alpha),\alpha)} \, ,\label{eq:Jprop2}
\end{eqnarray}
where $K'(\alpha)=K(1-\alpha)$ and $K(\alpha)$ is the complete elliptic integral of the first kind, (\ref{eq:LameCN}) can be represented in terms of Weierstrass functions $\wp(\theta)$ by virtue of
\begin{eqnarray}
\wp(\theta~;g_2,g_3)&=&e_3+\frac{e_1-e_3}{\text{sn}^2(\sqrt{e_1-e_3}\theta,\alpha)} \;. \label{eq:Jprop3}
\end{eqnarray}
The freedom of choice of the prefactor of $\wp(z~;g_2,g_3)$ allows us to write the evolution equation as
\begin{eqnarray}
\left[\partial_\theta^2+p^2-2\wp(\theta+iK'(\alpha);~g_2,g_3)\right]f_p(\theta)=0 \, ,
\label{eq:eq45}
\end{eqnarray}
where the parameters for $\alpha=1/2$ are summarized in Tab.~\ref{tab:WeierstrassParam}. By expressing the momentum $p^2$ implicitly in terms of the Weierstrass function $\wp(z)$ with
\begin{eqnarray}
\wp(z~;g_2,g_3)=-p^2 \, ,
\label{eq:fppMap}
\end{eqnarray}
(\ref{eq:eq45}) reduces to the well known Lam\'{e} equation in the Weierstrass form. The independent solutions $U_p(\theta)$ and $U_p(-\theta)$ are given in terms of Weierstrass functions $\zeta(\theta;g_2,g_3)$ and $\sigma(\theta;g_2,g_3)$ and read \cite{Ince,Boyanovski}
\begin{eqnarray}
U_p(\theta)=e^{-\theta \zeta(z)}\frac{\sigma(\theta+z+iK'(\alpha))~\sigma(i K'(\alpha))}{\sigma(\theta+iK'(\alpha))~\sigma(z+iK'(\alpha))} \, .
\label{eq:LSol}
\end{eqnarray}
\begin{table}[t]
\begin{center}
\begin{tabular}{||c|c|c|c|c|c||}
\hline \hline
  $\Delta$ & $e_1$ & $e_2$ & $e_3$ & $g_2$ & $g_3$ \\ \hline
    1	&     1/2  &  0    &  -1/2 &  1    &  0  \\ \hline \hline
\end{tabular}\\
\caption{ \label{tab:WeierstrassParam} Parameters of Weierstrass Elliptic functions for $\alpha=1/2$}
\end{center}
\end{table}
To further discuss the properties of these solutions for different momenta $p$ one needs to investigate the mapping (\ref{eq:fppMap}). The Weierstrass elliptic function $\wp(z)$  takes real values on the sides of the fundamental rectangle. Furthermore as $p^2>0$ we can restrict to negative values of $\wp(z)$. The solution to (\ref{eq:fppMap}) can be parametrized as
\begin{eqnarray}
z&=&i\beta \qquad \qquad \qquad \text{for}~p^2>1/2 \label{eq:stable_band} \\
z&=&iK'(1/2)+\beta \quad~ \text{for}~0\leq p^2\leq1/2 \label{eq:resonance_band}
\end{eqnarray}
where $0<\beta<K(1/2)$. The growth-rates can be obtained from a Floquet analysis, i.e.\ by investigating
\begin{eqnarray}
U_p(\theta+2K(1/2))=e^{2K(1/2)F(p)}~U_p(\theta) \, .
\end{eqnarray}
From (\ref{eq:LSol}) and using properties of the Weierstrass functions \cite{AbramovichStegun} one finds
\begin{eqnarray}
F(p)=\frac{1}{K(1/2)}\left(z \zeta(K(1/2))-K(1/2)\zeta(z)\right) \, .
\label{eq:FloquetIndex}
\end{eqnarray}
If this has a non-vanishing real part then the solution exhibits exponential growth and belongs to the resonance band. This is the case for all modes with $p^2\leq1/2$. If in contrast $p^2>1/2$, the Floquet index $F(p)$ is purely imaginary. Then the mode exhibits stable oscillations and therefore belongs to the stable band. In particular, the stable modes do not exhibit growth and are thus not relevant relevant for our discussion of nonlinear effects of the evolution. We therefore concentrate on the properties of the resonance band.

\subsection*{B.1 Resonance Band}
\begin{figure}[t]
\begin{minipage}[t]{\linewidth}
\centering
\includegraphics[width=\linewidth]{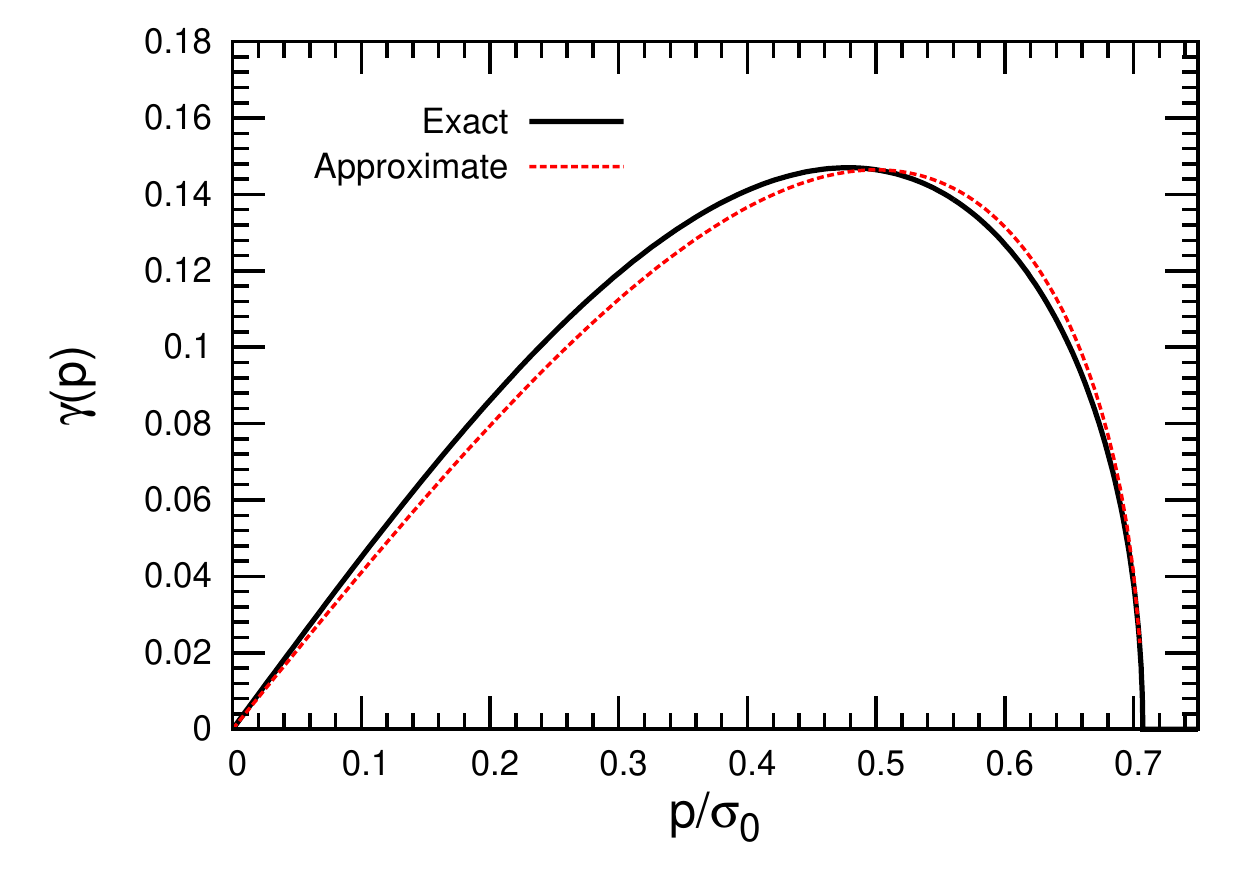}
\end{minipage}
\begin{minipage}[b]{\linewidth}
\centering
\includegraphics[width=\linewidth]{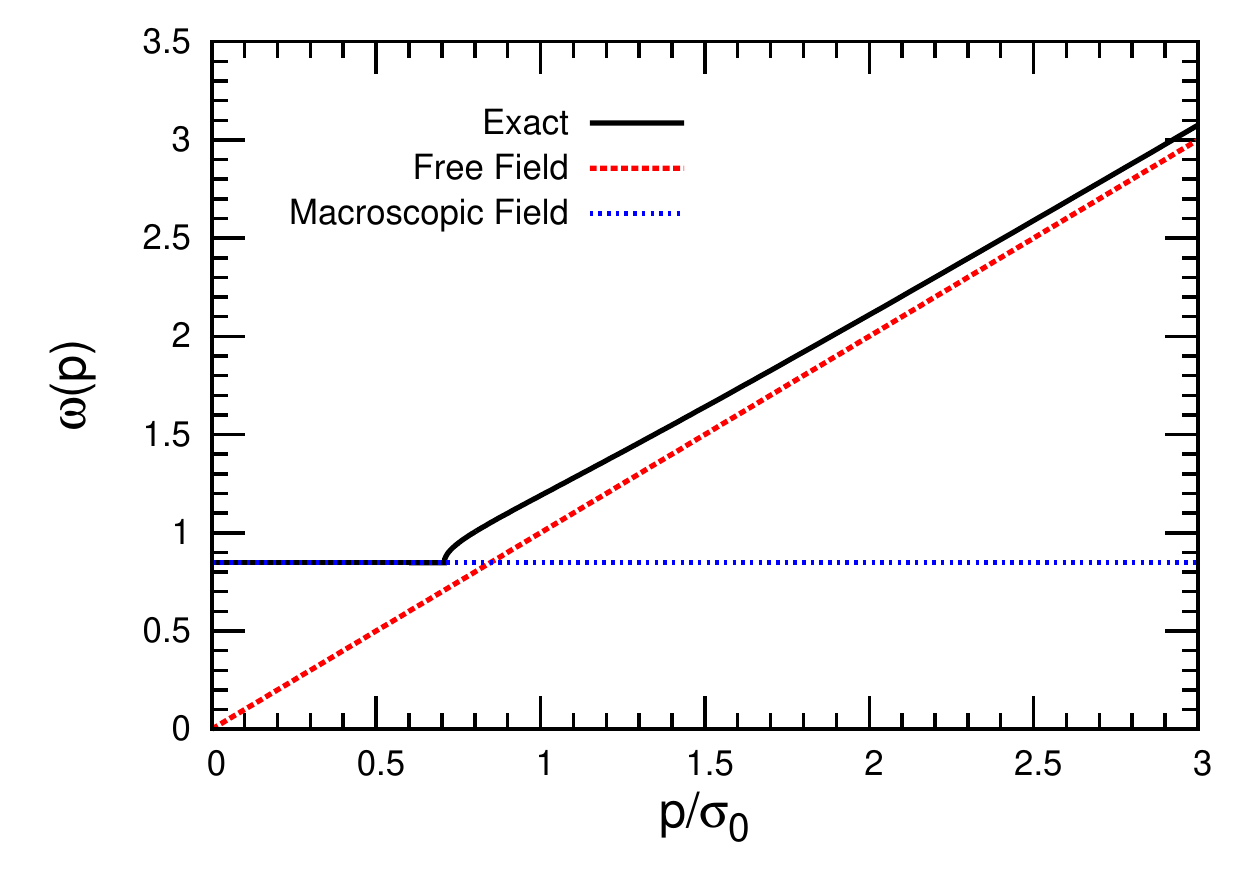}
\end{minipage}
\caption{ \label{fig:grExactVsApprox} (color online) Properties of the solution of the Lam\'{e} equation in Minkowski space-time. (\textbf{left}) Comparison of growth rates of fluctuations from the exact solution of (\ref{eq:FloquetIndex}) (black) with the approximation of (\ref{eq:Z_app_root}) (red dashed). (\textbf{right}) Oscillation frequency obtained from the exact solution (\ref{eq:FloquetIndex}) of the Lam\'{e} equation (black). Also shown is the free field limit (red dashed) and the oscillation frequency of the macroscopic field (blue dashed).}
\end{figure}

To further study the resonance band it is useful to invert the defining equation
\begin{eqnarray}
z(p)=iK'\left(\frac{1}{2}\right)+\beta(p)\;.
\end{eqnarray}
By use of the identities (\ref{eq:Jprop1})-(\ref{eq:Jprop3}) and performing the steps (\ref{eq:Jprop1})-(\ref{eq:fppMap}) 'backwards' we find
\begin{eqnarray}
p^{2}=\frac{1}{2}\textrm{cn}^{2}\left(\beta(p),\frac{1}{2}\right)
\label{eq:betaRes}
\end{eqnarray}
for resonant modes satisfying the inequality (\ref{eq:resonance_band}). To obtain a more intuitive expression for the growth-rates the Floquet indices $F\left(p\right)$ can be expressed as
\begin{eqnarray}
Z(u;~\alpha)=\frac{\pi}{2K\left(\alpha\right)}\frac{\vartheta_{4}'\left(\frac{\pi u}{2K\left(\alpha\right)}\right)}{\vartheta_{4}\left(\frac{\pi u}{2K\left(\alpha\right)}\right)} \;,
\end{eqnarray}
where $\vartheta_{j}(u)$, $j=1,\;2,\;3,\;4$ are the Jacobi theta functions. By use of the relations between Weierstrass zeta and Jacobi theta functions
\begin{eqnarray}
\zeta(z) & = & \frac{\zeta(K\left(\alpha\right))z}{K\left(\alpha\right)}+\frac{\pi}{2K\left(\alpha\right)}\frac{\vartheta_{1}'\left(\frac{\pi z}{2K\left(\alpha\right)}\right)}{\vartheta_{1}\left(\frac{\pi z}{2K\left(\alpha\right)}\right)}\label{eq:zeta} \;,
\end{eqnarray}
and using 
\begin{eqnarray}
\frac{\vartheta_{1}'\left(u+\frac{\pi}{2}\frac{K'(\alpha)}{K(\alpha)}\right)}{\vartheta_{1}\left(u+\frac{\pi}{2}\frac{K'(\alpha)}{K(\alpha)}\right)} & = & \left(\frac{\vartheta_{4}'(u)}{\vartheta_{4}(u)}-i\right)\label{eq:theta1} \;,
\end{eqnarray}
it is straightforward to obtain the final result
\begin{eqnarray}
F\left(p\right) & = & i\frac{\pi}{2K\left(1/2\right)}-Z(\beta(p);~1/2)\label{eq:unstable} \;.
\end{eqnarray}
The Jacobi zeta function $Z(\beta;~1/2)$ is purely real and therefore yields the growth-rate
\begin{eqnarray}
\gamma(p)=Z(\beta(p);~1/2)\,,\quad \beta(p)=\text{cn}^{-1}(\sqrt{2p^2}) \, .
\label{eq:gammaExact}
\end{eqnarray}
The growth rate as a function of momentum is shown in the left panel of Fig.~\ref{fig:grExactVsApprox}. The imaginary part in (\ref{eq:unstable}) is precisely the oscillation frequency of the macroscopic field. It is useful to apply approximations to the growth rate in (\ref{eq:gammaExact}) in order to obtain the explicit momentum dependence. To simplify the expression (\ref{eq:gammaExact}) we expand the inverse Jacobi cosine according to
\begin{equation}
\text{cn}^{-1}(z;~\alpha)=\frac{2K(\alpha)}{\pi}\cos^{-1}\left(z\right)+\mathcal{O}(\alpha)\;.
\end{equation}
In the next step we expand the Jacobi zeta function in a $q$-series
\begin{equation}
Z\left(\beta,\,1/2\right)=\frac{2\pi}{K\left(1/2\right)}\sum_{n=1}^{\infty}\frac{q^{n}}{1-q^{2n}}\sin\left(\frac{2\pi n\beta}{2K\left(1/2\right)}\right)
\end{equation}
where $q=e^{-\pi}$. By keeping the leading term in the above $q$-series and using trigonometric identities we obtain the final result
\begin{equation}
\gamma(p)\simeq\frac{4\pi e^{-\pi}}{K\left(1/2\right)}\sqrt{2p^{2}\left(1-2p^{2}\right)}\;.\label{eq:Z_app_root}
\end{equation}
This is compared to the exact growth-rate in Fig.~\ref{fig:grExactVsApprox}. While the overall behavior is described rather well, one observes that there are some deviations for the maximally amplified modes. For the latter one finds in the above approximation
\begin{eqnarray}
\gamma_{0}\simeq\frac{2\pi~e^{-\pi}}{K\left(1/2\right)}\;,\qquad \text{for} \qquad p_{0}\simeq\frac{1}{2}.
\end{eqnarray} 

\section*{Appendix C: Quasi-static approximation}

In the quasi-static approximation discussed in Sec.~ \ref{sec:instexp} we assume that the time dependence of the momenta is slow on the time scale of one period of oscillation of the macroscopic field. The explicit time dependence of the momenta $\tilde{p}(\theta)$ in the evolution equation for the fluctuations
\begin{eqnarray}
\left[\partial_\theta^2+\tilde{p}^2(\theta)+\text{cn}^2\left(\theta-\theta_0;~1/2\right)\right]&&F_{\bot}(\theta,\theta',p_T,\nu)=0 \nonumber \\
\label{eq:expLame}
\end{eqnarray}
can therefore be neglected when inferring the momentum dependent growth-rate. In practice, this corresponds to replacing $p\rightarrow \tilde{p}(p_T,\nu,\theta)$ for all expressions presented in appendix B. We test this approximation below, where we present a comparison between numerical simulations of  (\ref{eq:expLame}) and the quasi-static approximation.

\subsection*{C.1 Set-in and freeze-out times in the quasi-static approximation}

\begin{figure}[t]
\centering
\includegraphics[scale=0.6]{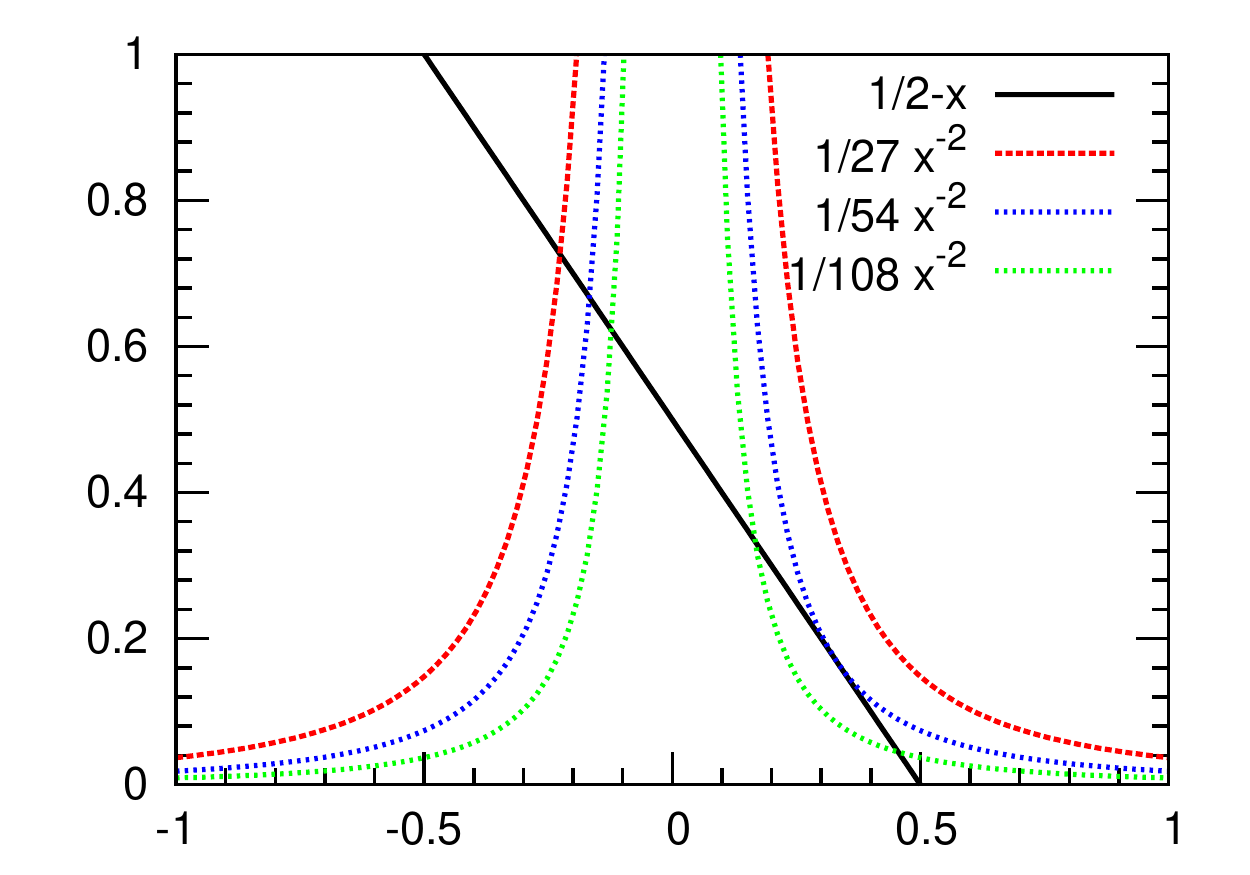}
\caption{ (color online) \label{fig:VisRoots} Visualization of (\ref{eq:GrowthTimeEqRes}) for different values of $p_T^4\nu^2/(\sigma_0^4(\sigma_0\tau_0)^2)$. There is always one negative solution which is irrelevant for the initial value problem. If $p_T^4\nu^2/(\sigma_0^4(\sigma_0\tau_0)^2)$ is smaller than the critical value of $1/54$ there are two positive solutions corresponding to the set-in and freeze-out times of the primary instability.
}
\end{figure}

By replacing $p\rightarrow \tilde{p}(p_T,\nu,\theta)$ the condition for modes to be contained in the resonance band, becomes time dependent. For $\tilde{p}^2(\theta)=2\theta/(3\sigma_0\tau_0)~ p_T^2/\sigma_0^2+9\nu^2/(4\theta^2)$ the condition then reads
\begin{eqnarray}
\frac{2\theta}{3(\sigma_0\tau_0)}~\frac{p_T^2}{\sigma_0^2}+\frac{9}{4}\frac{\nu^2}{\theta^2}\leq\frac{1}{2}\;. 
\label{eq:GrowthTimeEq}
\end{eqnarray}
For the initial value problem we are only interested in the solutions in the forward light-cone, i.e. solutions for which $\theta>0$. Before considering the generic case $p_T\neq0$ and $\nu\neq0$ we consider briefly the special cases where one of the two vanishes. For vanishing transverse momentum the condition is satisfied if
\begin{eqnarray}
\theta\geq\frac{\sqrt{18}}{2}\, \nu\;, 
\end{eqnarray}
which suggests that exponential growth sets in with a delay. In contrast for vanishing $\nu$ growth is limited to the time when
\begin{eqnarray}
\theta\leq(\sigma_0\tau_0)\frac{3}{4}\frac{\sigma_0^2}{p_T^2}\;, 
\end{eqnarray}
suggesting that exponential growth stops for later times. For the generic case (\ref{eq:GrowthTimeEq}) can be solved graphically. First we rescale $\theta'/\theta =2/(3 \sigma_0\tau_0)~p_T^2/\sigma_0^2$ yielding
\begin{eqnarray}
\frac{\nu^2p_T^4}{\sigma_0^4(\sigma_0\tau_0)^2}\frac{1}{\theta'^2}\leq\frac{1}{2}-\theta' \;.
\label{eq:GrowthTimeEqRes} 
\end{eqnarray}
This suggests that the existence of an unstable window only depends on the value of $\nu^2p_T^4$, whereas the precise time also depends on $p_T^2$. The existence of real solutions is visualized in 
Fig.~\ref{fig:VisRoots}, where we show the LHS and the RHS of (\ref{eq:GrowthTimeEqRes}) for different values of $\nu^2p_T^4$. One observes that one solution is always real and negative and therefore not relevant for the study of the initial value problem. The other solutions are real for $\nu^2p_T^4<\sigma_0^4(\sigma_0\tau_0)^2/54$. In this case the solutions are given by
\begin{eqnarray}
\theta'_{Start}&=&\frac{1}{6}\left[1-e^{-i\pi/3}/\alpha-e^{i\pi/3}\alpha\right] \;, \\
\theta'_{End}&=&\frac{1}{6}\left[1+1/\alpha+\alpha\right] \;,\\
\alpha&=&\left(1-2\beta+\sqrt{4\beta~\left(\beta-1\right)}\right)^{1/3} \;,\\
\beta&=&54~\frac{p_T^4\nu^2}{\sigma_0^4(\sigma_0\tau_0)^2}\;.
\end{eqnarray}
The solution is real if and only if $|\alpha|=1$. Indeed this is the case for resonant modes, i.e.\ modes with $\beta<1$. We can exploit this fact to rewrite the solution for the starting  and ending time as
\begin{eqnarray}
\theta'_{Start}&=&\frac{1}{6}\left(1+2\sin(\phi/3-\pi/6)\right) \\
\theta'_{End}&=&\frac{1}{6}\left(1+2\cos(\phi/3)\right) \\
\phi&=&2~\text{arctan}\left(\sqrt{ \frac{1}{\beta^{-1}-1}}\right)
\end{eqnarray}
which is manifestly real. We note that $\phi \in (0,\pi)$ for modes satisfying (\ref{eq:GrowthTimeEqRes}) and therefore $\theta'_{Start}\leq\theta'_{End}$ as suggested by the naming.

\subsection*{C.2 Tests of the approximation}
\begin{figure}
\begin{minipage}[t]{\linewidth}
\centering
\includegraphics[width=\linewidth]{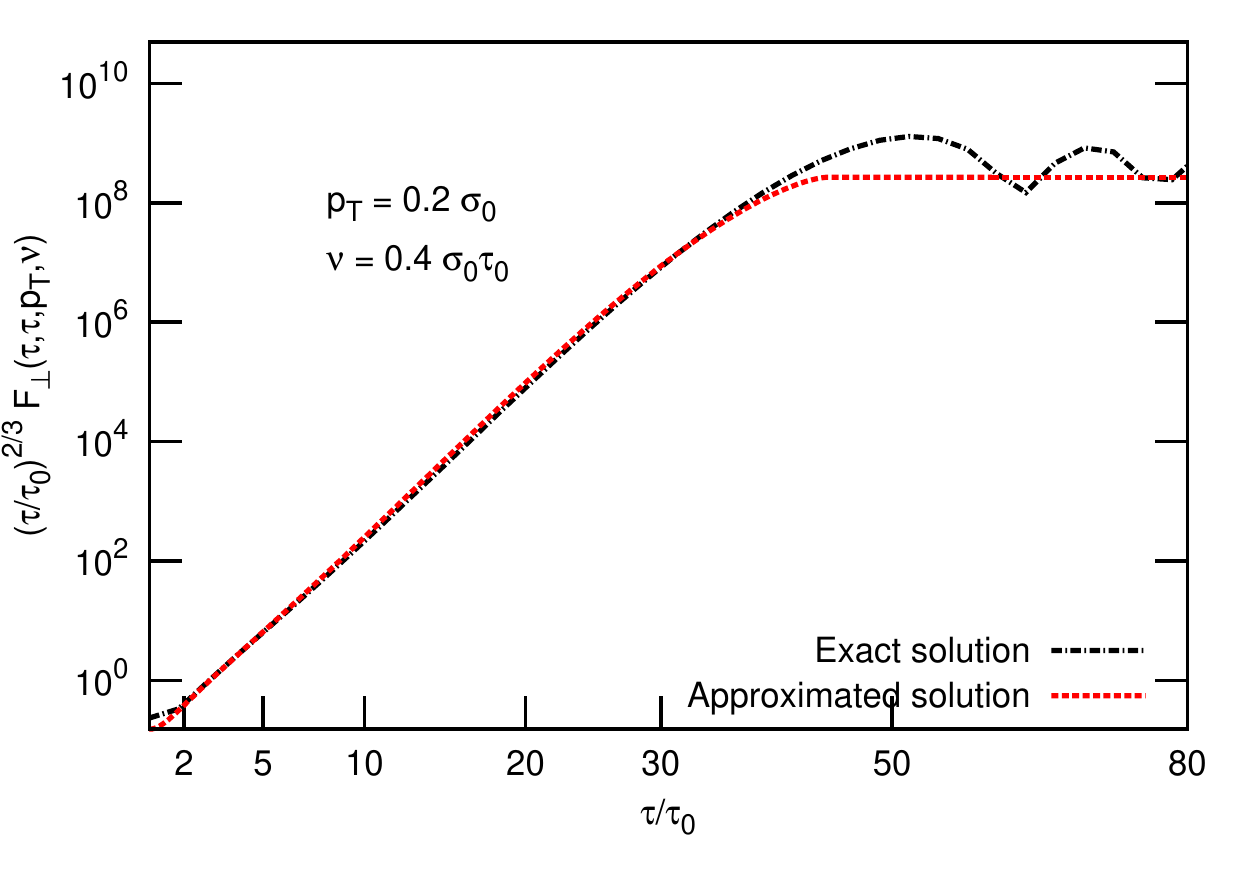}
\end{minipage}
\begin{minipage}[b]{\linewidth}
\centering
\includegraphics[width=\linewidth]{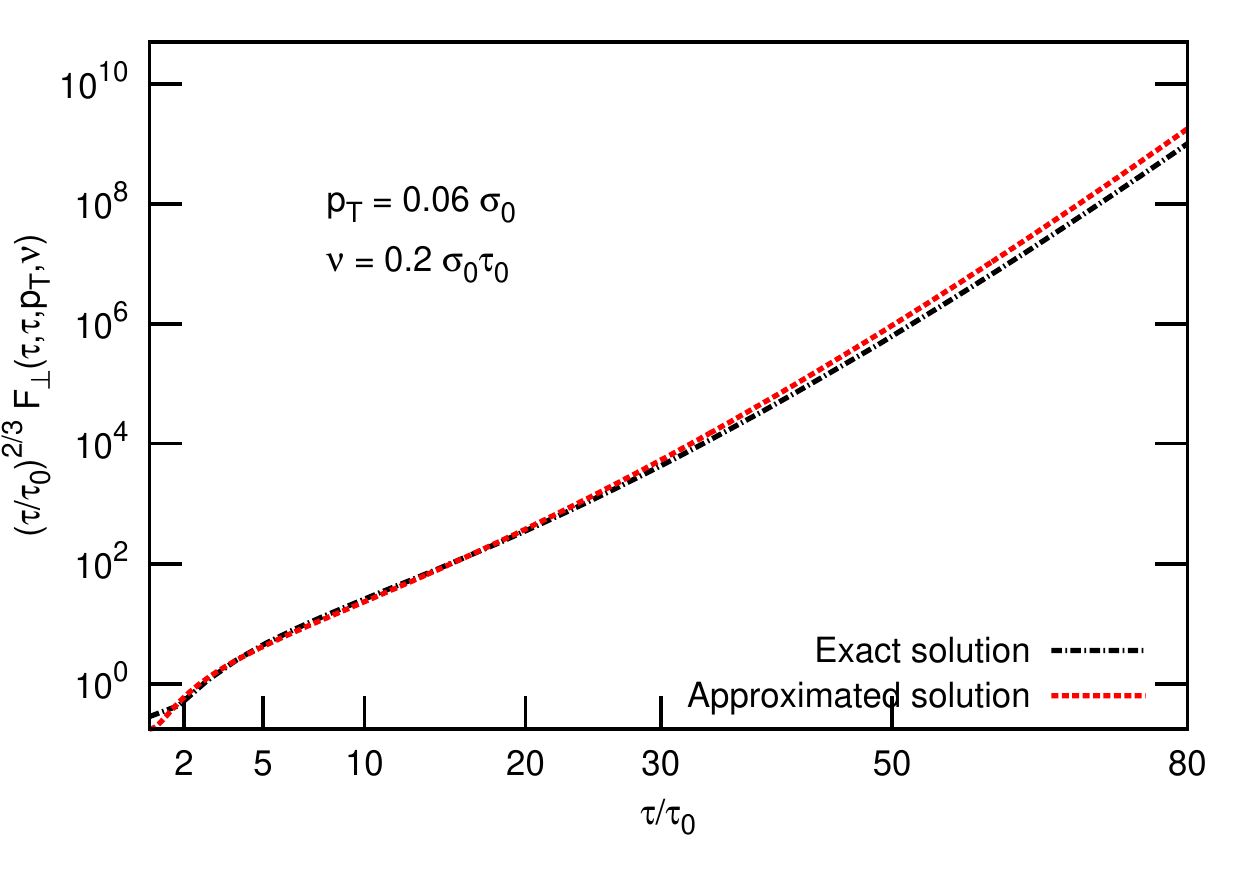}
\end{minipage}
\caption{\label{fig:QSA-Test}(color online) Time evolution of the transverse fluctuations $F_{\perp}\left(\tau,\tau,p_{T},\nu\right)$ averaged over one period of oscillation of the macroscopic field.
The black lines correspond to the numerical solution of the linearized evolution equations and red lines show the corresponding solutions in the quasi-static approximation where the growth rates
have been integrated in time numerically. The momenta are chosen as (\textbf{top}) $p_{T}=0.2\;\sigma_{0}$,
$\nu=0.4\;\sigma_{0}\tau_{0}$ and (\textbf{bottom}) $p_{T}=0.06\;\sigma_{0}$,
$\nu=0.2\;\sigma_{0}\tau_{0}$. The time axis is scaled as
$\left(\tau/\tau_{0}\right)^{2/3}$.
}
\end{figure}

The quasi-static approximation yields growth rates which can be
integrated in time numerically. In Fig.~\ref{fig:QSA-Test} this estimate is compared to the full time evolution of transverse fluctuations $F_{\bot}\left(\tau,\tau,p_{T},\nu\right)$. For times when the considered modes are outside the resonance band we assumed constant solutions in terms of the conformal variables. One observes that the evolution is indeed well described by the quasi-static approximation.

\section*{Appendix D: Initial conditions}
The initial conditions are formulated in terms of the conformal field variables as
\begin{eqnarray}
\tilde{\phi}(\theta_0)=\sqrt{\frac{6N}{\lambda}} \;, \qquad \partial_{\theta}\left.\tilde{\phi}(\theta)\right|_{\theta=\theta_0}=0\; \label{eq:ICfield}
\end{eqnarray}
for the macroscopic field. We choose the initial fluctuations to be exponentially suppressed at high momenta as
\begin{eqnarray}
\tilde{F}(\theta_0,\theta_0,p_T,\nu)&=&\frac{\sigma_0}{2\tilde{\omega}_p} e^{-\tilde{\omega}_p/\sigma_0}\, , \\
\partial_{\theta} \left.\tilde{F}(\theta,\theta',p_T,\nu)\right|_{\theta=\theta'=\theta_0}&=&
\partial_{\theta'} \left.\tilde{F}(\theta,\theta',p_T,\nu)\right|_{\theta=\theta'=\theta_0} \nonumber \\
&=&0 \, ,\\
\partial_{\theta} \partial_{\theta'} \left.\tilde{F}(\theta,\theta',p_T,\nu)\right|_{\theta=\theta'=\theta_0}&=&\frac{\tilde{\omega}_p}{2 \sigma_0} e^{-\tilde{\omega}_p/\sigma_0} \, ,
\label{eq:ICF}
\end{eqnarray}
where $\tilde{\omega}_{p\bot}=\sqrt{p_T^2+\nu^2/\tau_0^2+\sigma_0^2}$ for transverse modes and for  longitudinal modes we employ $\tilde{\omega}_{p\|}=\sqrt{p_T^2+\nu^2/\tau_0^2+3\sigma_0^2}$. Even though the choice of the initial conditions is somewhat arbitrary we find that the dependence on the initial conditions is rather weak as long as high momentum modes are sufficiently suppressed to avoid cut-off dependencies. In this situation the early-time dynamics is dominated by soft modes and the spectral shape is governed by the primary instability after a short period of time.  In terms of the original field variables and its derivatives with respect to proper time the initial conditions (\ref{eq:ICfield})-(\ref{eq:ICF}) translate to
\begin{eqnarray}
\phi(\tau_0)=\sigma_0\sqrt{\frac{6N}{\lambda}} \,, \quad \partial_{\tau}\left.\phi(\tau)\right|_{\tau=\tau_0}=-\frac{\sigma_0}{3\tau_0}\sqrt{\frac{6N}{\lambda}}
\end{eqnarray}
for the macroscopic field, where the first derivative originates from the time-dependent rescaling of the fields and accounts for the decay. Similarly for the fluctuations one finds
\begin{eqnarray}
F(\tau_0,\tau_0,p_T,\nu)&=&\frac{\sigma_0}{2\tilde{\omega}_p} e^{-\tilde{\omega}_p/\sigma_0} \, ,\\
\partial_{\tau} \left.F(\tau,\tau',p_T,\nu)\right|_{\tau=\tau'=\tau_0}&=&
\partial_{\tau'} \left.F(\tau,\tau',p_T,\nu)\right|_{\tau=\tau'=\tau_0}\nonumber \\
&=&-\frac{F(\tau_0,\tau_0,p_T,\nu)}{3\tau_0} \, ,\\
\partial_{\tau} \partial_{\tau'} \left.F(\theta,\theta',p_T,\nu)\right|_{\tau=\tau'=\tau_0}&=&\frac{F(\tau_0,\tau_0,p_T,\nu)}{9\tau_0^2} \nonumber \\
&+&\frac{\tilde{\omega}_p \sigma_0}{2} e^{-\tilde{\omega}_p/\sigma_0} \;.
\end{eqnarray}

\end{document}